\documentclass[aps,prc,superscriptaddress,twocolumn,10pt,longbibliography]{revtex4-2}
\usepackage{braket}
\usepackage{amsthm}
\usepackage{ascmac}
\usepackage{amssymb, amsmath}
\usepackage{bm}
\usepackage{physics}
\usepackage[dvipdfm]{graphicx}
\usepackage[dvipdfm]{color}
\usepackage{autobreak}
\usepackage{docmute}
\usepackage{times}
\usepackage[T1]{fontenc}
\usepackage{color}
\newcommand{\up}{\!\uparrow}
\newcommand{\down}{\!\downarrow}

\newcommand{\mev}{{\mathrm{MeV}}}
\newcommand{\fm}{{\mathrm{fm}}}
\newcommand{\zi}{\,{\mathrm{i}}}

\begin{document}
\title{
Phase transitions in the inner crust of neutron stars within the superfluid band theory:\\
Competition between $^1\text{S}_0$ pairing and spin polarization
under finite temperature and magnetic field
}


\author{Kenta Yoshimura}
\email[]{yoshimura.k.af21@m.isct.ac.jp}
\affiliation{Department of Physics, School of Science, Institute of Science Tokyo, Tokyo 152-8551, Japan}

\author{Kazuyuki Sekizawa}
\email[]{sekizawa@phys.sci.isct.ac.jp}
\affiliation{Department of Physics, School of Science, Institute of Science Tokyo, Tokyo 152-8551, Japan}
\affiliation{Nuclear Physics Division, Center for Computational Sciences, University of Tsukuba, Ibaraki 305-8577, Japan}
\affiliation{RIKEN Nishina Center, Saitama 351-0198, Japan}


\date{March 8, 2025}

\begin{abstract}
\begin{description}
\item[Background]
Phase transitions of matter under changes of external environment such as temperature and magnetic field have attracted great interests to various quantum many-body systems. Several phase transitions must have occurred in neutron stars as well such as transitions from normal to superfluid/superconducting phases and crust formation. While the temperature of a proto-neutron star is as high as 10\,MeV ($\approx 10^{11}$\,K) or higher, which are above critical temperatures for the emergence of superfluidity and crust formation, it cools rapidly down to 0.1\,keV ($\approx 10^6$\,K) already after hundreds of years. While ordinary neutron stars have surface magnetic field strength of around $10^{12}$\,G, those having higher magnetic field strength of $10^{14\text{--}15}$\,G or higher, so-called magnetars, have been observed. To uncover detailed evolution of neutron stars from their birth to later years, it is desired to develop fully microscopic approaches that take into account effects of superfluidity/superconductivity, finite temperature and magnetic field, on the same footing.


\item[Purpose]
The main purpose of this work is twofold: 1) to extend the formalism of the fully self-consistent superfluid nuclear band theory, developed in our previous work [K.~Yoshimura and K.~Sekizawa, Phys.\ Rev.\ C \textbf{109}, 065804 (2024)], for finite-temperature and finite-magnetic-field systems; 2) to explore possible phase transitions of nuclear matter by varying temperature and magnetic field.

\item[Methods]
We employ the superfluid band theory which is based on the Kohn-Sham density functional theory (DFT) for superfluid systems with a local treatment of paring, known as superfluid local density approximation (SLDA), subjected to the Bloch boundary conditions. We assume periodic spatial variation along $z$-direction with uniform distribution along $xy$-direction, allowing us to describe the slab phase as well as uniform nuclear matter. The finite-temperature extension is achieved in a similar manner as a finite-temperature Hartree-Fock-Bogoliubov calculation. Magnetic field effects are introduced taking into account both the Landau levels formation of relativistic electrons and the couplings of the magnetic field with nucleons' magnetic moments.

\item[Results]
We have performed superfluid band theory calculations for the slab phase of neutron star matter at $n_\text{B}=0.04$, 0.05, 0.06, and 0.07\,fm$^{-3}$ under various sets of temperature and magnetic field. From the results without magnetic field ($B=0$), we find that the superfluidity of neutrons disappears at around $k_\text{B}T=0.6$--$0.9\,\mev$, and ``melting'' of nuclear slabs, that is, a structural change into the uniform matter, takes place at around $k_\text{B}T=2.5$--$4.5\,\mev$. By turning on the magnetic field, we find that
protons' spin gets polarized at around $B=10^{16}$\,G, whereas neutrons' spin is kept unpolarized on average up to around $B=10^{17}$\,G. Intriguingly, our microscopic calculations reveal that neutrons' spin is actually polarized \textit{locally} inside and outside of the slab already at $B\sim10^{16}$\,G, while keeping the system unpolarized in total. We show that the local polarization of neutrons' spin is caused by an interplay of $^1\text{S}_0$ pairing among neutrons and spin-dependent interactions between neutron and protons.

\item[Conclusions]
We have demonstrated validity and usefulness of the fully self-consistent superfluid nuclear band theory for describing neutron star matter under arbitrary temperature and magnetic field. Critical temperatures and magnetic fields have been predicted for 1) superfluid to normal transition, 2) crust formation, and 3) spin polarization, under conditions relevant to realistic neutron star environments.

\end{description}
\end{abstract}

\maketitle

\section{INTRODUCTION}

How has each neutron star evolved from its birth to the present day? To answer this fundamental question, it is essential to understand the detailed properties of nuclear matter across a wide range of temperatures and magnetic field strengths. Among the theoretical frameworks available, nuclear density functional theory (DFT)~\cite{nakatsukasa2016a,Colò2020}—which encompasses both relativistic and non-relativistic mean-field approaches~\cite{bender2003}—stands out as one of the most powerful microscopic methods for describing nuclear matter as a many-nucleon quantum system.
Extensive efforts have been made to model the complex, non-uniform structures of nuclear matter, so-called ``pasta phases,'' within the inner crust of neutron stars~\cite{Ravenhall1983, Hashimoto1984}. These include both static~\cite{magierski2002,gogelein2008,newton2009,pais2012,grill2014,schuetrumpf2015b,fattoyev2017,schuetrumpf2019,schuetrumpf2020} and dynamic~\cite{schuetrumpf2013,schuetrumpf2014,schuetrumpf2015a} mean-field calculations. A more sophisticated and self-consistent description has been achieved by the nuclear DFT-based band theory, which was first applied to the slab phase~\cite{kashiwaba2019}, and later extended to include time-dependent phenomena~\cite{sekizawa2022}, superfluidity~\cite{yoshimura2024,almirante2024}, and even the rod phase~\cite{almirante2024a}.
Although it is, of course, sufficient to investigate structures of cold, ordinary neutron stars, it is insufficient to uncover the evolution from supernova matter through a hot proto-neutron star to a cold one. This article aims to establish the theoretical framework of the superfluid band theory at finite temperature and magnetic field. As a first step, we apply it to the slab and uniform phase of neutron star matter to demonstrate its feasibility.
The present work aims to establish a theoretical framework for the superfluid band theory that incorporates both finite-temperature and magnetic-field effects. As a first step, we demonstrate its feasibility by applying it to the slab and uniform phases of neutron star matter.

The nuclear band theory may not yet be a widely known in the nuclear physics community. The band theory of solids~\cite{ashcroft1976}, which is at the heart of solid-state physics, properly accounts for a periodic potential in quantum mechanical theories by imposing the Bloch boundary conditions. 
In the inner crust of neutron stars, free (dripped) neutrons permeate a crystalline lattice formed by nuclear clusters. To properly quantify the effects of this periodic potential on the dripped superfluid neutrons, it is essential to employ the framework based on the band theory. The first realistic calculations of such band structure effects in the inner crust were reported in 2005 for slab and rod phases~\cite{carter2005}, and for Coulomb lattices of spherical nuclei~\cite{chamel2005}.
Notably, it was shown that Bragg scattering of dripped neutrons off the periodic nuclear lattice can lead to a substantial reduction in the superfluid fraction~\cite{chamel2005,chamel2012,chamel2017}. This phenomenon, known as the ``entrainment effect,'' presents a significant challenge to standard models of pulsar glitches~\cite{andersson2012a,chamel2013,haskell2015a}.
To achieve a more conclusive understanding of entrainment, fully self-consistent band theory calculations based on nuclear density functional theory (DFT) have been developed. The first such self-consistent nuclear band theory calculation, which neglected pairing correlations, was carried out for the slab phase~\cite{kashiwaba2019}, and later extended to time-dependent scenarios~\cite{sekizawa2022}. Interestingly, these studies revealed that in the slab phase, the band structure induces the opposite behavior—an enhancement rather than a reduction of mobility—commonly referred to as the ``anti-entrainment effect''.
All of these calculations have been performed within the framework of band theory to determine the neutrons' effective mass.
On the other hand, there have also been many intriguing attempts to compute the superfluid fraction directly from the theory of superfluidity~\cite{watanabe2017, Watanabe2022, almirante2024, almirante2024a, almirante2025, Chamel2025}.
Although some of these studies incorporate the effects of the band structure, they still report a remaining entrainment effect, which is inconsistent with above-mentioned studies.
To systematically discuss such discrepancies and their relation to the underlying phenomena, it will be necessary to carry out comparative calculations in various systems, such as in two and three dimensions, and further developments in this direction are desirable.
More recently, we have extended the nuclear band theory framework to fully include both neutron superfluidity and proton superconductivity in a self-consistent manner~\cite{yoshimura2024}. These calculations confirm that the anti-entrainment effect persists in the slab phase even when neutron pairing correlations are included.

By further extending our theoretical framework to encompass systems at finite temperature and magnetic field, we aim to construct a fully self-consistent microscopic theory of neutron star matter. This unified approach will be applicable across the entire density range from the crust to the outer core, and under a wide variety of astrophysical conditions.
Naturally, a complete description of neutron star formation necessitates advanced simulation codes of core-collapse supernovae, which in turn require accurate microscopic inputs. Our goal is to provide such reliable microscopic information on the state of nuclear matter across the extreme conditions encountered during these explosive astrophysical events.

Concerning finite-temperature effects, the temperature during and immediately after a supernova explosion can reach 10~MeV or higher—sufficient to drastically alter the properties of nuclear matter. 
For example, such high temperatures can modify the equation of state of nuclear matter~\cite{Lassaut1987} and alter the energy structure within nuclei~\cite{bonche1984}.
Additionally, it is also believed that various phase transitions may occur within the extremely hot matter.
For instance, neutron superfluidity vanishes around $T \sim 1$~MeV, and the crust is expected to \textit{melt}, transitioning into uniform nuclear matter at several MeV. 
Such structural and compositional changes in the nuclear pasta phases impact neutrino scattering cross sections and, consequently, the opacity to neutrino flux~\cite{schuetrumpf2020}, as well as influence the cooling behavior of neutron stars~\cite{potekhin2015}.
Temperature also strongly affects the pairing properties in the inner crust, while the presence of nuclear clusters modifies the specific heat~\cite{sandulescu2004,pastore2012}. In addition, band structure effects modify the effective mass of unbound neutrons, thereby influencing thermal conductivity~\cite{chamel2013b}.
Moreover, neutron superfluidity not only suppresses neutrino emissivity~\cite{baiko2001,flowers1976}, but also introduces additional neutrino emission via pair breaking and formation (PBF) processes~\cite{flowers1976a,Voskresensky1987,leinson2009,leinson2010}. The quantitative evaluation of these contributions—especially their temperature and structure dependence—remains essential for modeling the thermal evolution of neutron stars.

On the other hand, the presence of strong magnetic fields in neutron stars can significantly alter the properties of nuclear matter through magnetic interactions with neutrons, protons, and electrons. Observations of \textit{magnetars} over the past decades~\cite{makishima2014, turolla2015, kaspi2017, esposito2021} suggest surface magnetic field strengths on the order of $10^{14\text{--}15}$\,G. At such magnitudes, the electron energy spectrum becomes discretized due to Landau quantization, a quantum mechanical effect that modifies the equation of state (EoS) of crustal matter and influences the nuclear composition in the outer crust~\cite{arteaga2011, chamel2012a, basilico2015, parmar2023a, sekizawa2023a}.
Theoretical studies further indicate that magnetic fields may be even stronger in the stellar interior, at least locally as large as $10^{17}$\,G, \textit{e.g.} in a form of a toroidal magnetic field~\cite{bonanno2003, naso2008, frieben2012}. 
Additionally, according to the Virial theorem and magnetohydrodynamics simulations, the upper limit on the neutron-star magnetic fields could be on the order of $B\approx 10^{18}$~G~\cite{potekhin1996, potekhin1999, broderick2000, Ventura:2001br}.
At such superstrong field strengths, shifts in single-particle energies of neutrons and protons become comparable to the MeV scale. Consequently, quantum shell structures, level ordering, and the deformation properties of nuclear clusters can be significantly modified~\cite{arteaga2011, basilico2015, stein2016, jiang2024}.
Moreover, recent calculations have proposed that under extreme magnetic fields approaching $10^{18}$\,G, superheavy nuclei—including elements beyond the current nuclear chart—could emerge as energetically stable constituents of the outer crust~\cite{sekizawa2023a, basilico2024}. Such findings suggest the possibility of exotic nuclear configurations in the magnetized environment of neutron stars.

We point out here that the interplay between pairing correlations and spin polarization in the presence of a magnetic field is intriguing and important. A superstrong magnetic field exceeding $B \sim 10^{17}$G can, on one hand, break spin-singlet Cooper pairs, but on the other hand, may assist the formation of spin-triplet Cooper pairs (see, \textit{e.g.}, Ref.~\cite{tajima2023}), which may alter thermal conductivity and the cooling rate. When a magnetic field breaks Cooper pairs, neutrons and protons may become spin-polarized. It has been shown that the interplay between spin polarization and pairing correlations could manifest a unique pairing phase, known as the Larkin-Ovchinnikov-Fulde-Ferrell (LOFF) phase~\cite{fulde1964, larkin1964, casalbuoni2004}. Recently, effects of spin polarization in strongly correlated Fermionic systems, known as unitary Fermi gas, have been explored within superfluid (TD)DFT, predicting the possible existence of a spin-polarized droplet, dubbed ``ferron'' \cite{magierski2019,magierski2021}, and complex spin-polarized structural patterns~\cite{tuzemen2023}. Moreover, spatial modulation of spin orientations can form topological objects such as Skyrmions~\cite{bogdanov2001,zhou2015,gobel2021}. A fully microscopic investigation of such exotic phases in nuclear matter has not yet been achieved to date, and this work is positioned as a first step towards exploring the above-mentioned exotic possibilities in the nuclear physics context.

In the present paper, we extend the theoretical framework of the self-consistent superfluid band theory \cite{yoshimura2024} for systems at finite-temperature and in finite magnetic fields. The theory is based on the Kohn-Sham DFT and its superfluid extension employing a local treatment of pairing, known as superfluid local density approximation (SLDA)~\cite{bulgac2002, bulgac2002a, jin2021}. The finite-temperature extension is achieved in the same manner as finite-temperature Hartree-Fock-Bogoliubov calculations~\cite{goodman1981, duguet2020}. The magnetic field effects are incorporated in both the magnetic interaction acting on the magnetic moments of neutrons and protons (see, \textit{e.g.}, Refs.~\cite{stein2016, jiang2024}) and the formation of the Landau levels of relativistic electrons (see, \textit{e.g.}, Ref.~\cite{chamel2012a}). 
Considering the possibility of extremely strong magnetic fields induced by effects such as the aforementioned toroidal magnetic fields, we investigate field strengths up to the theoretical limit of the order of $10^{17}$~G.
By applying the extended framework to the slab phase of neutron star matter, we explore the states of nuclear matter under various sets of temperature and magnetic fields. Intriguingly, we find nontrivial local polarizations of neutron spin at $B\simeq10^{16}$\,G, which is arising from an interplay between $^1\text{S}_0$ pairing correlations among neutrons and spin-dependent interactions between neutrons and protons.

The article is organized as follows. In Sec.~\ref{Sec:Formulation}, the theoretical framework of the fully self-consistent superfluid band theory is described, especially focusing on the extensions to finite-temperature and finite-magnetic-field systems. In Sec.~\ref{Sec:Results}, the results of numerical calculations are presented, showing how nuclear matter properties are altered with varying temperature and magnetic field. A summary and a future prospect are given in Sec.~\ref{Sec:Summary}.

\section{FORMULATION}

\subsection{Self-consistent nuclear band theory for superfluid systems} \label{Sec:Formulation}

In this section, we provide the formulation of the fully self-consistent band theory based on nuclear DFT for superfluid systems. The fully self-consistent band theory was first achieved for the slab phase in 2019 \cite{kashiwaba2019}, which was extended for time-dependent phenomena in 2022 \cite{sekizawa2022}. To take into account neutron superfluidity, we have recently extended \cite{yoshimura2024} the theoretical framework based on a superfluid DFT, known as SLDA.

Writing down explicitly spin ($\sigma=\;\up,\,\down$) and isospin ($q=n,p$) degrees of freedom, the coordinate-space representation of the HFB equation is given by
\begin{widetext}
\begin{equation}
    \mqty(\hat{h}_{\up\up}^{(q)}(\bm{r})-\lambda & \hat{h}_{\up\down}^{(q)}(\bm{r}) & 0 & \Delta(\bm{r}) \\
    \hat{h}_{\down\up}^{(q)}(\bm{r}) & \hat{h}_{\down\down}^{(q)}(\bm{r})-\lambda & -\Delta(\bm{r}) & 0\\
    0 & -\Delta^*(\bm{r}) & -\hat{h}_{\up\up}^{(q)*}(\bm{r}) + \lambda & -\hat{h}_{\up\down}^{(q)*}(\bm{r})\\
    \Delta^*(\bm{r}) & 0 & -\hat{h}^{(q)*}_{\down\up}(\bm{r}) & -\hat{h}^{(q)*}_{\down\down}(\bm{r})+\lambda)
    \mqty({u}^{(q)}_{\mu}(\bm{r}\up)\\
    {u}^{(q)}_{\mu}(\bm{r}\down)\\
    {v}^{(q)}_{\mu}(\bm{r}\up)\\
    {v}^{(q)}_{\mu}(\bm{r}\down)) = 
    E^{(q)}_{\mu}\mqty({u}^{(q)}_{\mu}(\bm{r}\up)\\
    {u}^{(q)}_{\mu}(\bm{r}\down)\\
    {v}^{(q)}_{\mu}(\bm{r}\up)\\
    {v}^{(q)}_{\mu}(\bm{r}\down)),
\end{equation}
\end{widetext}
where we call $u_\mu^{(q)}(\bm{r}\sigma)$ and $v_\mu^{(q)}(\bm{r}\sigma)$ the quasiparticle wave functions, $\lambda$ is the chemical potential and $\Delta$ is the pairing field.
For the pairing field in this case we consider the local and spin-singlet pairing, $\Delta(\bm{r},\bm{r}^\prime)=\Delta(\bm{r})\equiv\Delta_{\up\down}(\bm{r})=-\Delta_{\down\up}(\bm{r})$. 

Within the superfluid band theory, we impose the Bloch boundary condition to the quasiparticle wave functions:
\begin{eqnarray}
    u^{(q)}_{\mu\bm{k}}(\bm{r}\sigma) &=& \frac{1}{\sqrt{\mathcal{V}}} \tilde{u}^{(q)}_{\mu\bm{k}}(\bm{r}\sigma)e^{\zi\bm{k}\cdot\bm{r}}\\
    v^{(q)}_{\mu\bm{k}}(\bm{r}\sigma) &=& \frac{1}{\sqrt{\mathcal{V}}} \tilde{v}^{(q)}_{\mu\bm{k}}(\bm{r}\sigma)e^{\zi\bm{k}\cdot\bm{r}},
\end{eqnarray}
where $\mathcal{V}$ is a normalization volume and the transformed Bloch wave functions satisfy
\begin{equation}
    \begin{aligned}
        \tilde{u}^{(q)}_{\mu\bm{k}}(\bm{r}+\bm{T},\sigma) &= \tilde{u}^{(q)}_{\mu\bm{k}}(\bm{r}\sigma)\\
        \tilde{v}^{(q)}_{\mu\bm{k}}(\bm{r}+\bm{T},\sigma) &= \tilde{v}^{(q)}_{\mu\bm{k}}(\bm{r}\sigma),
    \end{aligned}
\end{equation}
with the imposed lattice vector $\bm{T}$.
The indexes $\mu$ and $\bm{k}$ are referred to as a band index and a Bloch wave number, respectively. The above definitions are consistent with imposing the Bloch boundary condition to the single-particle wave functions \cite{yoshimura2024}. We can derive the HFB equation for the dimensionless functions, $\tilde{u}_{\mu\bm{k}}^{(q)}(\bm{r}\sigma)$ and $\tilde{v}_{\mu\bm{k}}^{(q)}(\bm{r}\sigma)$ \cite{yoshimura2024}, and the resulting equations read:
\begin{widetext}
\begin{eqnarray}
    \mqty(\hat{h}_{\up\up}^{(q)}+\hat{h}_{\bm{k},\up\up}^{(q)}-\lambda & \hat{h}_{\up\down}^{(q)}+\hat{h}_{\bm{k},\up\down}^{(q)} & 0 & \Delta_q \\
    \hat{h}_{\down\up}^{(q)}+\hat{h}_{\bm{k},\down\up}^{(q)} & \hat{h}_{\down\down}^{(q)}+\hat{h}_{\bm{k},\down\down}^{(q)}-\lambda & -\Delta_q & 0\\
    0 & -\Delta_q^* & -\hat{h}_{\up\up}^{(q)*}-\hat{h}_{-\bm{k},\up\up}^{(q)*} + \lambda & -\hat{h}_{\up\down}^{(q)*}-\hat{h}_{-\bm{k},\up\down}^{(q)*}\\
    \Delta_q^* & 0 & -\hat{h}^{(q)*}_{\down\up}-\hat{h}^{(q)*}_{-\bm{k},\down\up} & -\hat{h}^{(q)*}_{\down\down}-\hat{h}^{(q)*}_{-\bm{k},\down\down}+\lambda)
    \mqty(\tilde{u}^{(q)}_{\mu\bm{k}}(\bm{r}\up)\\
    \tilde{u}^{(q)}_{\mu\bm{k}}(\bm{r}\down)\\
    \tilde{v}^{(q)}_{\mu\bm{k}}(\bm{r}\up)\\
    \tilde{v}^{(q)}_{\mu\bm{k}}(\bm{r}\down))
    = 
    E^{(q)}_{\mu\bm{k}}\mqty(\tilde{u}^{(q)}_{\mu\bm{k}}(\bm{r}\up)\\
    \tilde{u}^{(q)}_{\mu\bm{k}}(\bm{r}\down)\\
    \tilde{v}^{(q)}_{\mu\bm{k}}(\bm{r}\up)\\
    \tilde{v}^{(q)}_{\mu\bm{k}}(\bm{r}\down)).
\end{eqnarray}
\end{widetext}
Here we omit the coordinate index $(\bm{r})$ in the HFB matrix for a concise expression. The single-particle Hamiltonian with the Bloch wave number $\hat{h}_{\bm{k},\sigma\sigma'}$ can be formally obtained by the following replacement of the derivative operator in $\hat{h}_{\sigma\sigma'}$:
\begin{equation}
    \grad \to \grad + \zi\bm{k}.
\end{equation}
For more detailed description, we refer readers to Refs.~\cite{sekizawa2022,yoshimura2024}.

\subsection{Energy density functional}

For practical applications, we employ nuclear energy density functional (EDF) approach. In the present work, we use a Skyrme-type EDF, as in our previous work \cite{yoshimura2024}. We work with the nuclear EDF of the following form:
\begin{equation}
\frac{E_\text{nucl}}{N_B} = \frac{1}{N_B}\int \bigl(
\mathcal{E}_\text{kin}(\bm{r}) + \mathcal{E}_\text{Sky}(\bm{r}) + \mathcal{E}_\text{Coul}^{(p)}(\bm{r}) + \mathcal{E}_\text{pair}(\bm{r})
\bigr) \dd \bm{r},
\end{equation}
with the baryon number $N_B$.
The kinetic energy part, $\mathcal{E}_\text{kin}$, the nuclear interaction part, $\mathcal{E}_\text{Sky}$, the Coulomb part, $\mathcal{E}_\text{Coul}$, and the pairing part, $\mathcal{E}_\text{pair}(\bm{r})$, in the nulcear EDF are given, respectively, by
\begin{eqnarray}
\mathcal{E}_\text{kin}(\bm{r}) &=& \sum_{q=n,p}\frac{\hbar^2}{2m_q}\tau_q(\bm{r}),\\
\mathcal{E}_\text{Sky}(\bm{r}) &=& \mathcal{E}^\text{even}_\text{Sky}(\bm{r}) + \mathcal{E}^\text{odd}_\text{Sky}(\bm{r})\\
\mathcal{E}_\text{Coul}^{(p)}(\bm{r}) &=& \frac{1}{2}V_\text{Coul}(\bm{r})n_p(\bm{r})
-\frac{3e^2}{4}\biggl(\frac{3}{\pi}\biggr)^{1/3}n_p^{4/3}(\bm{r}),\\
\mathcal{E}_\text{pair}(\bm{r}) &=&  -\sum_{q=n,p} \Delta_q(\bm{r})\kappa_q^*(\bm{r}),
\end{eqnarray}
where $\hbar$ is the reduced Plank's constant, $m_q$ denotes the mass of a neutron ($q=n$) and a proton ($q=p$), and $e$ is the elementary charge. 
The interaction part is devided into the time-even and time-odd contributions, which are written as
\begin{eqnarray}
    \mathcal{E}^\text{even}_\text{Sky}(\bm{r}) &=& \sum_{t=0,1}\Bigl[
C_t^\rho[n_0]n_t^2(\bm{r}) + C_t^{\Delta\rho}n_t(\bm{r})\partial_z^2n_t(\bm{r}) \nonumber\\[-1.5mm]
&&\hspace{6mm}+ C_t^\tau\bigl(n_t(\bm{r})\tau_t(\bm{r})-J_t^2(\bm{r})\bigr)
\Bigr]\\[2mm]
    \mathcal{E}^\text{odd}_\text{Sky}(\bm{r}) &=& \sum_{t=0,1}\Bigl[ C_t^{\bm{s}}[n_0]\bm{s}_t^2(\bm{r}) + C_t^{\Delta\bm{s}}\bm{s}_t(\bm{r})\bm{\cdot}\Delta\bm{s}_t(\bm{r}) \nonumber\\
&&\hspace{6mm}+ C_t^{\bm{T}}\bigl(\bm{s}_t(\bm{r})\bm{\cdot T}_t(\bm{r})-\bm{j}_t^2(\bm{r})\bigr)\Bigr]. \nonumber\\[2mm]
\end{eqnarray}
In these formulae $n$, $\tau$, $\bm{j}$, $\bm{s}$, $\bm{T}$, and $J$ are various nucleonic densities, whose explicit definitions will be given later. The index $t$ represents isoscalar ($t=0$, \textit{e.g.} $n_0=n_n+n_p$) and isovector ($t=1$, \textit{e.g.} $n_1=n_n-n_p$) components. $C_t^\text{X}$ ($\text{X}=\rho,\Delta\rho,\dots$) are the parameters of the functional, which are determined to reproduce the known properties of finite nuclei and nuclear matter. In the functional shown above, the spin-orbit term is omitted, because it does not play any role in the systems with spatial modulations along a certain single dimension, like a slab phase. 
Although the time-odd components vanish in the static cases with time-reversal symmetry, in the present work we include them in the functional, as the external magnetic field explicitly breaks the symmetry. 
Note, however, that since the $\boldsymbol{s\cdot}\Delta\bm{s}$ term is known to cause a spin instability, we set $C_t^{\Delta\bm{s}}=0$ throughout our analysis~(see, \textit{e.g.}, Refs.~\cite{Hellemans2012,Sekizawa2013}).

For the electron part, we adopt an EDF for relativistic electron gas,
\begin{equation}
\frac{E_\text{elec}}{N_B} = \frac{1}{N_B}\bigl(\mathcal{E}_\text{kin}^{(e)} + \mathcal{E}_\text{Coul}^{(e)}\bigr)a,
\end{equation}
where
\begin{eqnarray}
\mathcal{E}_\text{elec} &=& \int_0^{p_\text{F}}\frac{4\pi p^2\dd p}{(2\pi)^3}\sqrt{m_\text{e}^2c^4+p^2c^2}\nonumber\\
&=& \frac{m_\text{e}^4c^5}{32\pi^2\hbar^3}(\sinh\theta_\text{F}-4\theta_\text{F}),\\
\mathcal{E}_\text{Coul}^{(e)} &=& \frac{3e^2}{8}\biggl(\frac{3}{\pi}\biggr)^{1/3}n_\text{e}^{4/3},
\end{eqnarray}
with $p_\text{F}=\hbar(3\pi^2n_\text{e})^{1/3}$ denoting the Fermi momentum with the electron number density $n_\text{e}$. $\theta_\text{F}$ is defined through the relation, $\varepsilon_\text{e}=\sqrt{m^2c^4+p_\text{F}^2c^2}=m_\text{e}c^2\cosh{\theta_\text{F}}$.
The Coulomb exchange term is evaluated with the Slater approximation. neutrality condition.
In a normal way the Coulomb potential and electron density are calculated via the Poisson equation
\begin{equation}
    \frac{\dd}{\dd z}V_{\mathrm{Coul}}(z) = -\frac{e^2}{\varepsilon_0}n_{\mathrm{ch}}(z),
\end{equation}
with $n_{\mathrm{ch}}(z)=n_p(z) - n_e$, and the charge neutrality condition
\begin{equation}
    \frac{1}{L_z}\int\, n_p(z)\dd z - n_e = 0.
\end{equation}
From this charge neutrality we find that the direct term of the electrons’ Coulomb energy vanishes, which has been already pointed out in Ref.~\cite{kashiwaba2019}.
It is to mention here that the electronic EDF will be modified in the presence of an external magnetic field, which will be discussed in Sec.~\ref{Sec:B-extension}.

From appropriate functional derivatives, one can derive the ordinary ($\bm{k}$-independent) single-particle Hamiltonian, divided into the time-even and odd components as the functional, written as
\begin{equation}
    \hat{h}^{(q)}(\bm{r}) = \hat{h}^{(q)}_\text{even} + \hat{h}^{(q)}_\text{odd},
\end{equation}
where each component is given by
\begin{eqnarray}
\hat{h}_{\sigma\sigma^\prime,\text{even}}^{(q)}(\bm{r}) &=& -\bm{\nabla}\cdot M^{(q)}(\bm{r}) \bm{\nabla} + U^{(q)}(\bm{r}), \\[2mm]
\hat{h}_{\sigma\sigma^\prime,\text{odd}}^{(q)}(\bm{r}) &=&-\bm{\nabla}\cdot \qty(\bm{\Lambda}^{(q)}(\bm{r})\cdot\bm{\sigma}) \bm{\nabla}+\bm{\Sigma}^{(q)}(\bm{r})\bm{\cdot\sigma} \nonumber\\[2mm]
&&\hspace{6mm}+ \frac{1}{2i}\qty[ \div \bm{I}^{(q)}(\bm{r}) + \bm{I}^{(q)}(\bm{r})\bm{\cdot}\grad ],
\nonumber\\[2mm]
\end{eqnarray}
as well as the $\bm{k}$-dependent one,
\begin{eqnarray}
\hat{h}_{\text{even},\bm{k}}^{(q)}(\bm{r}) &=& M^{(q)}(\bm{r})\bm{k}^2+ \hbar\bm{k\cdot}\hat{\bm{v}}_\text{even}^{(q)}(\bm{r}),\\[2mm]
\hat{h}_{\text{odd},\bm{k}}^{(q)}(\bm{r}) &=&  \qty(\Lambda^{(q)}(\bm{r})\cdot\bm{\sigma})\bm{k}^2 + \hbar\bm{k\cdot}\hat{\bm{v}}_\text{odd}^{(q)}(\bm{r}),
\end{eqnarray}
where
\begin{align}
    \hat{\bm{v}}^{(q)}_\text{even}(\bm{r}) &= -\mathrm{i}\hbar \Bigl[ \grad M^{(q)}(\bm{r}) + M^{(q)}(\bm{r})\grad  \Bigr]\\[2mm]
    \hat{\bm{v}}^{(q)}_\text{odd}(\bm{r}) &= -\mathrm{i}\hbar\Bigl[ \grad \qty( \bm{\Lambda}^{(q)}(\bm{r})\cdot\bm{\sigma} ) + \qty( \bm{\Lambda}^{(q)}(\bm{r})\cdot\bm{\sigma} )\grad\Bigr]\nonumber\\[2mm] & \hspace{8mm}+\frac{1}{\hbar} \bm{I}^{(q)}(\bm{r}).
\end{align}
The various mean-field potentials in the single-particle Hamiltonian are defined as follows:
\begin{eqnarray}
M^{(q)}(\bm{r}) &=& \frac{\hbar^2}{2m_q} + \sum_{q'=n,p}C_{q'}^{\tau(q)}n_{q'}(\bm{r}),\\
\bm{\Lambda}^{(q)}(\bm{r}) &=& \sum_{q^\prime=n,p} C^{\bm{T}(q)}_{q^\prime} \bm{s}_{q'}(\bm{r}),\\
    U^{(q)}(\bm{r}) &=& \sum_{q^\prime = n, p}\Big[ 2C^{\rho(q)}_{q^\prime}n_{q^\prime}(\bm{r}) + 2C^{\nabla\rho(q)}_{q^\prime}\partial^2_z n_{q^\prime}(\bm{r}) \nonumber\\
    &&\hspace{10mm} + C^{\tau(q)}_{q^\prime}\tau_{q^\prime}(\bm{r}) + 2n_0^\alpha(\bm{r})C^{\rho(q)}_{q^\prime D}n_{q^\prime}(\bm{r}) \Big]\nonumber\\[1mm]
    &&+ \alpha n_0^{\alpha-1}(\bm{r})\sum_{t = 0,1}C^\rho_{t}[n_0]n^2_t(\bm{r})\nonumber\\
    &&+ U_{\mathrm{Coul}}(\bm{r})\delta_{qp}
      + \sum_{q^\prime = n,p} \frac{\partial g_{q'\!,\text{eff}}}{\partial n_q}|\kappa_{q^\prime}(\bm{r})|^2,\\
    \bm{\Sigma}^{(q)}(\bm{r}) &=& \sum_{q^\prime=n,p}\Big[2C^{\bm{s}(q)}_{q^\prime}[n_0]\bm{s}_{q^\prime}(\bm{r})+ C^{\bm{T}(q)}_{q^\prime}[n_0]\bm{T}_{q^\prime}(\bm{r})\Big],\nonumber\\[-2mm]\\[1mm]
    \bm{I}^{(q)}(\bm{r}) &=& -2 \sum_{q^\prime = n, p}C^{\tau(q)}_{q^\prime}\bm{j}_{q^\prime}(\bm{r}).
\end{eqnarray}
In the above formulas, the coefficients with two isospin indices $C^{\mathrm{X}(q)}_{q^\prime}$ are the shorthand notations, defined by:
\begin{eqnarray}
    C_n^{\text{X}(q)} &\equiv& C_0^\text{X} + \eta_q C_1^\text{X},\\[1.5mm]
    C_p^{\text{X}(q)} &\equiv& C_0^\text{X} - \eta_q C_1^\text{X},
\end{eqnarray}
where X stands for the superscript of the coefficients, \textit{e.g.}, $\rho$, $\tau$, etc., and $\eta_q=+1$ ($-1$) for neutrons (protons).

In the SLDA formalism, the pairing field $\Delta_q(\bm{r})$ is local in space:
\begin{equation}
\Delta_q(\bm{r}) = -g_{q,\text{eff}}(\bm{r})\kappa_q(\bm{r}).
\end{equation}
Here, $g_{q,\text{eff}}$ is an effective pairing coupling constant calculated within the scheme of the superfluid local density approximation~\cite{bulgac2002,jin2021}.
With the natural energy cutoff $\hbar^2/2m (\pi/\Delta z)^2$, this is given by
\begin{equation}
\frac{1}{g_{q,\text{eff}}} = \frac{1}{g_0} - \frac{K}{8\pi^2M^{(q)}}\frac{\pi}{\Delta z},
\end{equation}
where $g_0$ denotes the bare coupling constant, $K$ is a numerical constant~\cite{jin2021} written as
\begin{equation}
    K = \frac{12}{\pi}\int_0^{\pi/4} \dd\theta \ln(1+1/\cos^2\theta) = 2.4427496... 
\end{equation}
and $\Delta z$ is the spatial mesh spacing. 
In this work we employ $g_0 =- 200\,\mathrm{MeV}\,\mathrm{fm}^3$ and $\Delta z = 0.5\,\mathrm{fm}$.
The anomalous density $\kappa_q$ is defined by
\begin{equation}
\kappa_q(\bm{r}) = \sum_{\mu\bm{k}} v^{(q)*}_{\mu\bm{k}\,\up}(\bm{r}\sigma)u^{(q)}_{\mu\bm{k}\,\down}(\bm{r}\sigma).
\end{equation}
We note that there is an additional contribution to $U^{(q)}(\bm{r})$ arising from the density dependence of the effective pairing coupling constant, which is given by
\begin{equation}
\pdv{g_{q'\!,\text{eff}}}{n_q}
= \bigl[g_{q'\!,\mathrm{eff}}(\bm{r})\bigr]^2\frac{K}{8\pi\Delta z}
\Bigl(\! M^{(q^\prime)}(\bm{r})\!\Bigr)^{\!\!\!-2}\!\!C^{\tau(q^\prime)}_{q}.
\end{equation}

The single-particle Hamiltonian are given as a functional of various densities. Those densities are given in terms of the quasiparticle wave functions as follows:
\begin{eqnarray}
    n_q(\bm{r}) &=& \sum_{\mu\bm{k}\sigma}\abs{v^{(q)}_{\mu\bm{k}}(\bm{r}\sigma)}^2,\\
    \tau_q(\bm{r}) &=& \sum_{\mu\bm{k}\sigma}\abs{\grad v^{(q)}_{\mu\bm{k}}(\bm{r}\sigma)}^2,\\
    \bm{j}_q(\bm{r}) &=& -\sum_{\mu\bm{k}\sigma}\mathrm{Im}\Bigl[v^{(q)*}_{\mu\bm{k}}(\bm{r}\sigma)\grad v_{\mu\bm{k}}(\bm{r}\sigma)\Bigr],\\
    \bm{s}_q(\bm{r}) &=& \sum_{\mu\bm{k}ss^\prime}v^{*(q)}_{\mu\bm{k}}(\bm{r}s)v^{(q)}_{\mu\bm{k}}(\bm{r}s^\prime)\bm{\sigma}_{ss^\prime},\\
    \bm{T}_q(\bm{r}) &=& \sum_{\mu\bm{k}ss^\prime}\qty[\grad v^{*(q)}_{\mu\sigma}(\bm{r}s)\cdot \grad v^{(q)}_{\mu\sigma}(\bm{r}s^\prime)]\bm{\sigma}_{ss^\prime},\\
    J_{q,\alpha\beta}(\bm{r}) &=& -\frac{1}{2i}\sum_{\mu\bm{k}ss^\prime} \Big[v^*_{\mu\bm{k}}(\bm{r}s^\prime)\bigl(\nabla_\alpha v_{\mu\bm{k}}(\bm{r}s)\bigr) \nonumber \\
    &&\hspace{13.5mm} - v^{(q)}_{\mu\bm{k}}(\bm{r}s)\bigl(\nabla_\alpha v^{*(q)}_{\mu\bm{k}}(\bm{r}s^\prime)\bigr)\Big]\qty[\sigma_\beta]_{ss^\prime},\label{Eq:J_mn}\nonumber\\
\end{eqnarray}
where $\alpha$ and $\beta$ in Eq.~\eqref{Eq:J_mn} are spatial indexes, $\alpha,\beta=x,y,z$.
In the case of the slab phase, the dimensionless Bloch wave functions depend only on $z$ coordinate perpendicular to the slabs, \textit{i.e.} $\tilde{u}_{\mu\bm{k}}^{(q)}(z\sigma)$ and $\tilde{v}_{\mu\bm{k}}^{(q)}(z\sigma)$ \cite{yoshimura2024}, which enables us to significantly reduce the computational cost. For more specific representations of densities and potentials for the slab phase, we refer the readers to Ref.~\cite{sekizawa2022,yoshimura2024}.

\subsection{Finite-temperature extension}

Next, let us introduce the finite-temperature extension of the superfluid band theory.
It is achieved by starting the from the grand canonical ensemble, whose application to the HFB theory is given in Ref.~\cite{goodman1981}.
Within the grand canonical ensemble, the thermodynamic equilibrium is represented as
\begin{equation}
    \delta\Omega = 0,
\end{equation}
where the grand potential $\Omega$ is given by
\begin{equation}
    \Omega = E - TS - \lambda N.
\end{equation}
In accordance with the equilibrium condition, the expectation value of physical quantities can be obtained with the following operator:
\begin{equation}
    \hat{\mathcal{D}} = \frac{1}{\mathcal{Z}}e^{-\beta(\hat{H} - \lambda \hat{N})},
\end{equation}
where $\beta=1/k_\text{B}T$ is the inverse temperature with the Boltzmann constant $k_\text{B}$ and $\mathcal{Z}$ denotes the partition function, $\mathcal{Z}=\Tr\qty[e^{-\beta(\hat{H}-\lambda \hat{N})}]$.

Applying the operator $\hat{\mathcal{D}}$ to the HFB theory, we can obtain the representations of the density matrix and the pairing tensor, respectively, as
\begin{eqnarray}
    \rho &=& V^*(1-f)V^{\mathrm{T}} + UfU^\dag,\\[1mm]
    \kappa &=& V^*(1-f){U}^{\mathrm{T}} + UfV^\dag,
\end{eqnarray}
where the matrix $f$ is composed of matrix elements $f_{\mu\nu}=f_\beta(E_\mu)\delta_{\mu\nu}$ with $f_\beta(E)$ the Fermi-Dirac distribution function,
\begin{equation}
    f_\beta(E) = \frac{1}{1+e^{\beta E}}.
\end{equation}
As a consequence, various densities are replaced by mixture of contributions from $u$- and $v$-components of the quasiparticle wave functions.
In general, we define the matrix vector $\bm{\rho} = [\rho, \tau, ...]$ and the calculation formula for each element as
\begin{equation}
    \bm{\rho} = \sum_{\mu\bm{k}\sigma} \bm{\mathcal{F}}[v^{(q)}_{\mu\bm{k}\sigma}],
\end{equation}
where $\bm{\mathcal{F}} = [\mathcal{F}_\rho, \mathcal{F}_\tau, \ldots]$ is a shorthand notation which, for instance, works as $\mathcal{F}_{\rho}[X] = \abs{X}^2$ for the number density, $\mathcal{F}_{\tau}[X] = \abs{\grad X}^2$ for the kinetic energy density and similar for the others. Using this definition, densities other than the anomalous density in the finite temperature system can be written as
\begin{equation}
    \bm{\rho}(\bm{r},T) = \sum_{\mu\bm{k}\sigma} \qty{f_\beta(E_{\mu\bm{k}}) \bm{\mathcal{F}}[u^{(q)}_{\mu\bm{k}\sigma}] + f_\beta(-E_{\mu\bm{k}}) \bm{\mathcal{F}}[v^{(q)}_{\mu\bm{k}\sigma}]},
\end{equation}
and the anomalous density is given by
\begin{equation}
    \kappa_q(\bm{r}) = \sum_{\mu\bm{k}}[f_\beta(-E_{\mu\bm{k}}) - f_\beta(E_{\mu\bm{k}})]v^{(q)*}_{\mu\bm{k}\,\up}(\bm{r}\sigma)u^{(q)}_{\mu\bm{k}\,\down}(\bm{r}\sigma).
\end{equation}

For a finite temperature system, an equilibrium solution is obtained by minimizing the Helmholtz's energy,
\begin{equation}
    F = E - TS,
\end{equation}
where the one-particle entropy $S$ is given by
\begin{eqnarray}
    S &=& -k_\text{B}\sum_\mu \Bigl[f_\beta(E_\mu)\ln f_\beta(E_\mu) \nonumber\\[-1.5mm]
    &&\hspace{10mm}+ \bigl[1-f_\beta(E_\mu)\bigr]\ln \bigl[1-f_\beta(E_\mu)\bigr] \Bigr].
\end{eqnarray}
Since we can calculate the total energy as a function of temperature, the specific heat can be directly calculated as
\begin{equation}
    C_V(T) = \pdv{E(T)}{T}.
    \label{Eq:def_specific_heat}
\end{equation}
We will use the specific heat to characterize phase transitions.

\subsection{Extension for systems under a magnetic field}\label{Sec:B-extension}

In this section, we describe how to introduce a magnetic field $\bm{B}$ into the superfluid band theory at arbitrary temperature $T$. There are two major modifications: the first one is modification of single-particle energies through the coupling between nucleonic magnetic moments and $\bm{B}$; the second one is modification of electrons' energies because of the Landau-Rabi quantization.
There are several researches dealing with the latter point \textit{e.g.} Ref.~\cite{arteaga2011, chamel2012a, basilico2015}.
The former point is explained in Ref.~\cite{stein2016},

The modification of the nuclear part can be achieved by introducing an additional term to the single-particle Hamiltonian as
\begin{equation}
    \hat{h}_{q,\sigma\sigma'} = \hat{h}^{(0)}_{q,\sigma\sigma'} +  \hat{h}^{(B)}_{q,\sigma\sigma'},
\end{equation}
where $\hat{h}^{(0)}_{q,\sigma\sigma'}$ denotes the original single-particle Hamiltonian without magnetic field. The second term, $\hat{h}^{(B)}_{q,\sigma\sigma'}$, which represents the magnetic effects, is given by
\begin{equation}
    \hat{h}^{(B)}_{q,\sigma\sigma'} = -\qty(\bm{l}\delta_{q,p} + g_q\frac{\bm{\sigma}}{2}) \bm{\cdot}\tilde{\bm{B}}_q,\label{Eq:hB_normal}
\end{equation}
where $l$ is the dimensionless orbital angular momentum (\textit{i.e.} $\bm{L} = \hbar\bm{l}$), $\bm{\sigma}$ is the Pauli matrices, and $\tilde{\bm{B}}_q = (e\hbar/2m_q c) \bm{B}$. Here, $g_q$ is the g-factors of neutrons and protons, which are given by
\begin{equation}
    g_n = -3.826,\quad g_p = +5.585
\end{equation}
The first term in Eq.~\eqref{Eq:hB_normal} denotes the orbital contribution which couples only with protons, while the second one represents the coupling with nucleons' intrinsic spin. Note that the first term is absent in the slab phase under study.

It is customary to quantify the magnetic field strength relative to the critical value at which energy of the electron's cyclotron motion reaches their rest mass,
\begin{equation}
    B_\text{c} = \frac{m_\text{e}^2c^3}{e\hbar} \simeq 4.41\times 10^{13}\,\mathrm{G},
\end{equation}
denoting $B_\star \equiv B / B_\text{c}$. Using this notation, $\tilde{B}_q$ can be written as
\begin{equation}
    \tilde{\bm{B}}_{q} = \frac{m_\text{e}^2}{2m_q} \bm{B}_\star.
\end{equation}
In this study, the magnetic field is assumed to be oriented solely along the $z$-axis, \textit{i.e.}, perpendicular to the slab. Although it would be intriguing to explore effects of the magnetic-field directions on the structure and energy, it would require extension of the theoretical framework to two- and three-dimensional systems. In the present work, we focus on the nature of superfluid and structural phase transitions induced by the magnetic field perpendicular to the slabs, and leave exploration of such orientation effects as a future work.

The energy of Landau levels of relativistic electrons is given by
\begin{equation}
    e_\nu = \sqrt{c^2p_z^2 + m_\text{e}^2c^4(1+2\nu B_\star)},
\end{equation}
where $\nu$ is an index of the landau levels which is a non-negative integer, and $p_z$ is a momentum along the $z$ axis parallel to the magnetic field (perpendicular to the slabs). The energies of occupied states should be below the chemical potential, $e_\nu \leq \mu_\text{e}$, for existing $p_z>0$, which is equivalent to
\begin{equation}
    \nu \leq \frac{1}{2B_\star}\qty(\frac{\mu_\text{e}^2}{m_\text{e}^2c^4} - 1).\label{eq:2-C:nucond}
\end{equation}
Defining the maximum integer satisfying the above condition as $\nu_{\mathrm{max}}$,
the electron number density and energy density are written by
\begin{eqnarray}
    n_\text{e} &=& \frac{2B_\star}{(2\pi)^2\lambda_\text{e}^3} \sum_{\nu=0}^{\nu_{\mathrm{max}}} g_\nu x_\text{e}(\nu),\\
    \mathcal{E}_\text{e} &=& \frac{B_\star m_\text{e}c^2}{(2\pi)^2\lambda_\text{e}^3} \sum_{\nu=0}^{\nu_{\mathrm{max}}} g_\nu(1+2\nu B_\star) \psi_+ \qty[\frac{x_\text{e}(\nu)}{\sqrt{1+2\nu B_\star}}]\nonumber\\&& - n_\text{e}m_\text{e}c^2,
\end{eqnarray}
with
\begin{eqnarray}
    \psi_\pm(x) &=& x\sqrt{1+x^2} \pm \ln(x+\sqrt{1+x^2}),\\
     x_\text{e}(\nu) &=& \sqrt{\gamma_\text{e}^2 - 1 - 2\nu B_\star},\\
     \gamma_\text{e} &=& \frac{\mu_\text{e}}{m_\text{e}c^2},\\
    \lambda_\text{e} &=& \frac{\hbar}{m_\text{e}c}.
\end{eqnarray}
The chemical potential $\mu_\text{e}$ should be determined in such a way that it obeys the $\beta$-equilibrium condition, $\mu_n = \mu_p + \mu_\text{e}$, and the charge neutrality condition, $n_\text{e} = \frac{1}{L_z}\int n_p(\bm{r})\dd\bm{r} = N_p/L_z$.

At this point, one might wonder whether the Landau quantization could also influence energy levels of protons. Indeed, the strong magnetic field on the order of $10^{17\text{--}18}$\,G may cause noneligible effects on the motion of protons parallel to the slabs. It may affect the chemical potential of protons, which, in turn, may change the proton fraction. We note that such effects were investigated in Ref.~\cite{broderick2000} and a significant effect was found at $B_\star\approx10^5$ (i.e.\ $B\approx 4.41\times10^{18}$\,G), where a reduction of pressure resulted in softening of the equation of state. Nevertheless, in the present work, we neglect the effect of Landau quantization on protons and leave its investigation as future work.


Note that the magnetic field breaks the time-reversal symmetry. One should thus introduce the time-odd densities such as $\bm{s}_q$, $\bm{T}_q$, and $\bm{j}_q$. Further, if the external magnetic field is strong enough, one may expect that the nucleons are spin-polarized. To quantify the spin polarization, we define the local spin polarization as
\begin{equation}
    p_q(z) = \frac{n_{q,\up}(z) - n_{q,\down}(z)}{n_{q,\up}(z) + n_{q,\down}(z)}
    = \frac{s_{z,q}(z)}{n_q(z)},
\end{equation}
where in the last expression we use the $z$ component of the time-odd spin density, $s_{z,q}(z)$. We also define the total spin polarization of the system,
\begin{equation}
    P_{q} = \frac{N_{q,\up} - N_{q,\down}}{N_{q,\up} + N_{q,\down}},
\end{equation}
where $N_{q,\sigma} = \int n_{q,\sigma}(\bm{r})\dd\bm{r}$. Without the magnetic field, the system is unpolarized and at certain critical magnetic field strength one should observe spin polarization as a sort of phase transition.

\section{Results and discussion}\label{Sec:Results}

\subsection{Computational Settings}


We use our own computational code to perform self-consistent superfluid band theory calculations. The computational settings are the same as in our previous work~\cite{yoshimura2024}, including a discretization step size of $\Delta z=0.5\,\mathrm{fm}$ and the SLy4 energy density functional, unless stated otherwise.
Spatial derivatives are evaluated using a spectral method with fast Fourier transforms (FFTs). The Poisson equation for the Coulomb potential is also solved using the FFT algorithm.
Ideally, one should determine the optimal cell period by minimizing the total energy with respect to the distance between neighboring slabs. We denote this period as $L_z$. However, we find that the optimal slab period, which minimizes the energy, is somewhat ambiguous to determine accurately.
In addition, Ref.~\cite{almirante2024} has pointed out that the optimal cell size depends significantly on the choice of the energy density functional.
Since our main objective here is to investigate qualitative effects of finite temperature and magnetic fields on the phase structures, we fix the cell size to $a = 30\,\fm$ or $a = 40\,\fm$ for simplicity.
For comparisons across different densities, we use $n_\text{B} = 0.04$, $0.05$, $0.06$, and $0.07\,\fm^{-3}$, consistent with our previous work~\cite{yoshimura2024}.

All the calculations have been performed under $\beta$-equilibrium and charge neutrality conditions.  
Although one may ask a question whether such an equilibrium condition holds even at finite temperature or in the presence of a magnetic field, we expect that the static ground state remains in $\beta$ equilibrium. This is because $\beta$ equilibrium represents the balance among the chemical potentials of the reaction-involved particles, a balance that remains unchanged even in the presence of external fields. We would like to mention here that in the case of supernova matter one should modify the $\beta$ equilibrium condition to include effects of degenerated neutrinos.

We remark that the band structure effects have only a minor impact on the results discussed in the present paper. 
Therefore, in the following, we focus on the effects of superfluidity, finite temperature, and magnetic field.
Nevertheless, all the results presented below include the band structure effects.

\subsection{Finite-temperature effects}

In this section, we restrict the analysis to the $B$\,$=$\,0 case (without magnetic field), focusing on how the properties of nuclear matter are affected by finite-temperature effects. We will investigate effects of finite magnetic field in the next section.

\subsubsection{Pairing and structural phase transitions}

\begin{figure}[t]
    \centering\vspace{-11mm}
    \includegraphics[width=\columnwidth]{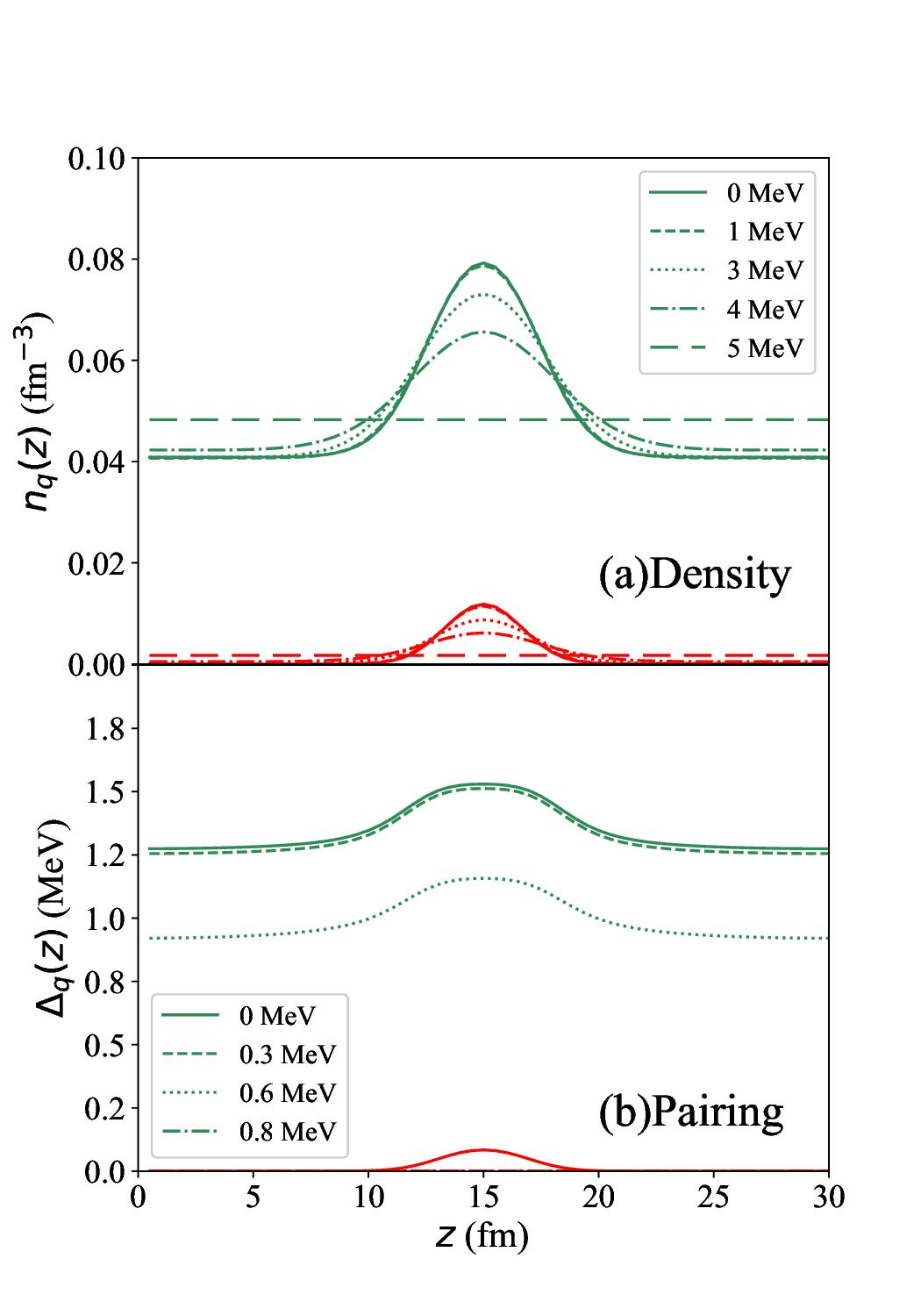}\vspace{-8mm}
    \caption{(a) Density distributions and (b) pairing fields of neutrons and protons are shown as a function of $z$ coordinate at four representative temperatures, $k_\text{B}T = 0$, $1$, $3$, $4$, and $5\,\mev$. 
    In both panels, the upper green lines indicate the distribution of neutrons' quantities, while the red lines are that for protons' ones.
    In ascending order of temperatures, solid, dashed, dotted, dash-dotted, and long-dashed lines are used.}
    \label{fig:FT_dens}
\end{figure}

\begin{figure}[t]
    \centering\vspace{-11mm}
    \includegraphics[width=\columnwidth]{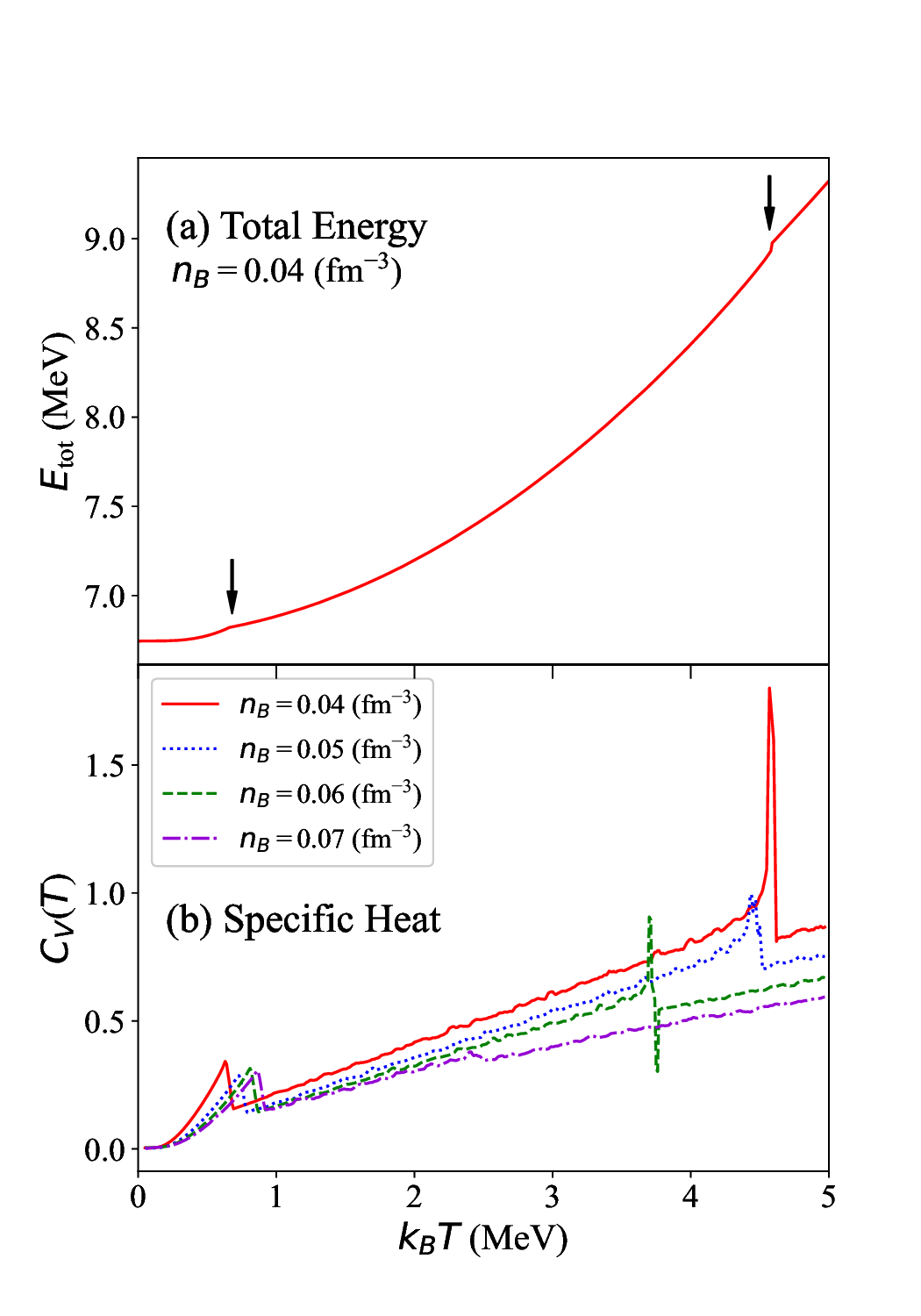}\vspace{-8mm}
    \caption{(a) Total energy per nucleon, $E_{\mathrm{tot}}$, is shown as a function of temperature for a fixed baryon density, $n_\text{B}=0.04\,\fm^{-3}$. Two arrows indicates the position of kinks implying phase transitions. 
    (b) Specific heat $C_V(T)$ is shown as a function of temperature, for different baryon number densities, $n_\text{B}=0.04$, 0.05, 0.06, and $0.07\,\fm^{-3}$.
    In ascending order of densities, solid, dashed, dotted, and dash-dotted lines are used.}
    \label{fig:energy_cvt}
\end{figure}

Figure~\ref{fig:FT_dens} shows spatial distributions of the nucleon number densities [Fig.~\ref{fig:FT_dens}(a)] and pairing fields [Fig.~\ref{fig:FT_dens}(b)] along the $z$ coordinate at various temperatures. The baryon number density and the slab period are fixed to $n_\text{B}=0.05\,\mathrm{fm}^{-3}$ and $a = 30\,\mathrm{fm}$, respectively. 
In panel (a), the nucleon number densities are shown for $k_\text{B}T=0$, $1$, $3$, $4$ and $5$ MeV.
In panel (b), the pairing fields are shown for $k_\text{B}T=0$, $0.3$, $0.5$, and $0.8$ MeV.
The solid, dashed, dotted, dash-dotted, and long-dashed lines correspond to increasing temperatures.
In both panels, green lines represent the neutron contributions, which have higher magnitudes, while red lines stand for the progon distributions, which are generally smaller.

From Fig.~\ref{fig:FT_dens}(a), we find that the density distributions remain almost unchanged up to $k_\text{B}T=1$~MeV. As the temperature increases further, the slab shape gradually becomes diffusive at $k_\text{B}T=3$\,MeV (dotted lines) and 4\,MeV (dash-dotted line), and eventually melts into uniform nuclear matter at $k_\text{B}T=5$~MeV (long-dashed lines) for both neutrons and protons. In contrast, Fig.~\ref{fig:FT_dens}(b) shows that the pairing field is highly sensitive to temperature. At $k_\text{B}T=0.3$~MeV, the proton pairing field vanishes, whereas neutron pairing field still maintains a sizable magnitude. As the temperature increases further, the neutron pairing field also vanishes as well at $k_\text{B}T = 0.8$~MeV.

These results indicate that the presence of two distinct phase transitions in nuclear matter as the temperature increases: 1) a pairing phase transition of neutrons (protons) from superfluid (superconducting) to the normal phase, and 2) a structural transition from nuclear pasta to uniform matter.
The former can be interpreted as a second-order phase transition, because of the observed smooth decrease in the pairing field. In contrast, since the structural transition entails a sudden change from the nonuniform phase to the gaseous homogeneous phase, the latter one is classified as a first-order transition.

\subsubsection{Heat capacity and critical temperatures}

The critical temperatures of the phase transitions can be determined more precisely by analyzing the specific heat, as defined in Eq.~\eqref{Eq:def_specific_heat}. 
To compute the specific heat, we perform finite-temperature calculations with a temperature increment of $k_\text{B}\Delta T = 0.01$~MeV, and evaluate the first derivative of the total energy using a 9-point finite-difference formula.
We have confirmed that the results are not sensitive to the choice of the order used in the finite-difference scheme.
Figures \ref{fig:energy_cvt}(a) and \ref{fig:energy_cvt}(b) show the total energy and the specific heat, respectively, calculated with a fixed slab period of $a = 30$,fm. In Fig.~\ref{fig:energy_cvt}(a), the total energy is plotted for a representative baryon number density of $n_\text{B} = 0.04\,\mathrm{fm}^{-3}$, while in Fig.~\ref{fig:energy_cvt}(b), the specific heat is shown for $n_\text{B} = 0.04$, $0.05$, $0.06$, and $0.07$~fm$^{-3}$.
From Fig.~\ref{fig:energy_cvt}(a), we observe that the total energy varies continuously but non-smoothly as a function of temperature, with noticeable kinks (\textit{i.e.}, abrupt changes in slope) at approximately $k_\text{B}T = 0.7$ and $4.5$~MeV (indicated by black arrows), the specific heat exhibits sharp peaks around the same temperatures.
We define the critical temperature of the phase transition as the temperature at which the specific heat reaches a local maximum, while disregarding minor thermal fluctuations that do not correspond to phase transitions. For $n_\text{B} = 0.04~\mathrm{fm}^{-3}$ (red solid line), the critical temperatures associated with the two peaks are found to be $0.63$ and $4.57~\mathrm{MeV}$, respectively.
Hereafter, we denote the lower critical temperature, corresponding to the superfluid phase transition, as $T_\text{c}^\text{sf}$, and the higher one, corresponding to the structural (melting) transition, as $T_\text{c}^\text{melt}$. As the baryon number density increases, we find that $T_\text{c}^\text{sf}$ shifts slightly to higher temperatures, whereas $T_\text{c}^\text{melt}$ tends to decrease.

In addition to the superfluid and structural (melting) phase transitions discussed above, another potentially important transition is the nuclear solid–liquid transition. This transition is generally associated with the loss of long-range crystalline order, giving rise to plasma-like uncorrelated motion of various nuclear species. Such behavior has been theoretically linked to the hydrodynamics of neutrons flowing around impurities, as discussed in Refs.~\cite{magierski2004, martin2016}.
Capturing this type of transition, however, lies beyond the scope of our present band-structure framework, as the method \textit{a priori} assumes a perfectly periodic system and cannot accommodate inherently unstable or disordered configurations.
Some efforts have previously been made to model the solid–liquid transition within the compressible liquid-drop model (CLDM) by approximately incorporating thermal kinetic effects~\cite{thi2023}. A similar strategy—based on introducing suitable correction terms—could, in principle, be adapted to extend our band-theoretic approach to account for such disorder-induced transitions.
Alternatively, one could model crystalline disorder more directly by enlarging the computational domain beyond a single period and explicitly including a long sequence of nuclear clusters. This approach would allow for a more realistic treatment of spatial fluctuations and disorder.
Nevertheless, we deliberately defer a detailed exploration of the solid–liquid transition to future work. In the present study, we focus exclusively on the superfluid and melting transitions, for which our method is most reliable and well-suited.


In Table~\ref{tab:critical_temperature}, we summarize the critical temperatures obtained for the baryon number densities for the two slab periods, $a =30$ and $40\,\fm$. The critical temperature is defined as the peak position of the kinks in the specific heat. 
The table clearly reveals systematic trends in the critical temperatures: as the baryon number density increases, the critical temperature for the pairing phase transition ($T_\text{c}^\text{sf}$) increases, whereas that for the structural (melting) transition ($T_\text{c}^\text{melt}$) decreases, irrespective of the slab period $a$.
This contrasting behavior can be understood as follows. An increase in baryon number density implies a greater number of nucleons per unit volume, which in turn requires higher thermal energy to break the pairing correlations. In contrast, the increasing diffuseness of the density distribution at higher densities leads to a configuration closer to uniform nuclear matter [\textit{cf.} Fig.~\ref{fig:FT_dens}(a)], thereby reducing the critical temperature associated with the structural transition.
It is also worth emphasizing that the critical temperatures exhibit only a weak dependence on the slab period $a$, thus validating our fixed-$a$ treatment adopted in the present study.

\begin{figure}[t]
    \centering\vspace{-11mm}
    \includegraphics[width=\columnwidth]{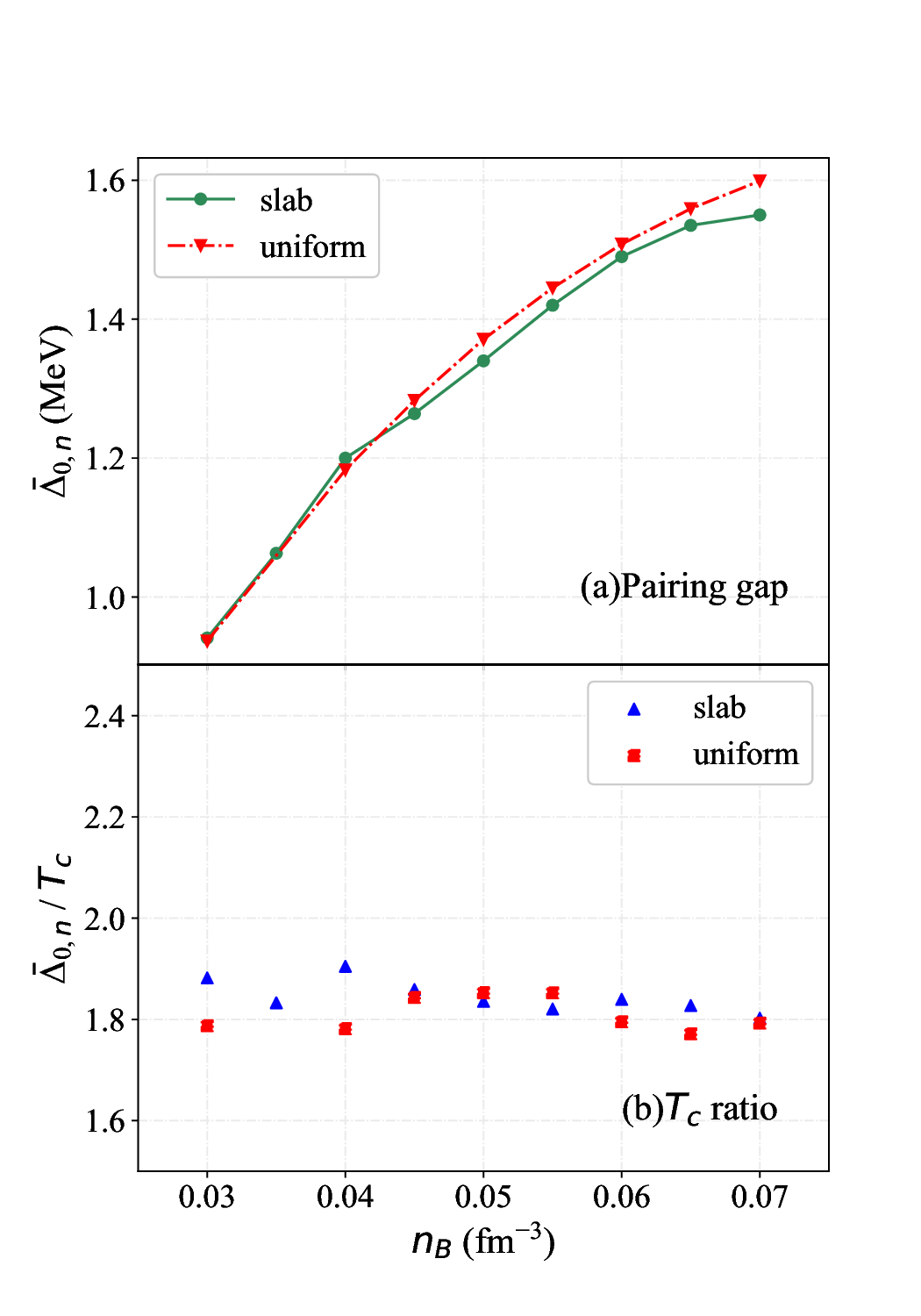}\vspace{-8mm}
    \caption{(a) The average pairing gap in the zero temperature $\Delta_0$, and (b) its ratio against the critical temperature $T_c$ plotted as a function of the baryon densities $n_B$. In both panels, results in the case of slab phases as well as uniform matter are demonstrated.}
    \label{fig:tcratio}
\end{figure}

\begin{table}[t]
    \centering
    \caption{Calculated critical temperatures for the superfluid phase transition, $T^{\mathrm{sf}}_c$, and the structural transition, $T^{\mathrm{melt}}_c$, are listed in units of MeV, for a couple of baryon number densities, $n_\text{B}=0.04$, 0.05, 0.06, and $0.07\,\fm^{-3}$, and the two slab periods, $a=30$ and $40\,\fm$.}\vspace{2mm}
    \begin{tabular*}{\columnwidth}{@{\extracolsep{\fill}}c|cccc}
        \hline\hline 
        $n_\text{B}\;(\fm^{-3})$ & 0.04 & 0.05 & 0.06 & 0.07 \\
        \hline
        $T_\text{c}^\text{sf}(a=30)$ (MeV) & 0.63 & 0.73 & 0.81 & 0.86\\
        $T_\text{c}^\text{melt}(a=30)$ (MeV)& 4.57 & 4.32 & 3.70 & 2.40\\
        \hline
        $T_\text{c}^\text{sf}(a=40)$ (MeV)& 0.65 & 0.75 & 0.83 & 0.89\\
        $T_\text{c}^\text{melt}(a=40)$ (MeV)& 4.82 & 4.44 & 3.44 & 2.75\\
        \hline\hline
    \end{tabular*}
    \label{tab:critical_temperature}
\end{table}
Figure~\ref{fig:tcratio} shows (a) the average neutron pairing gap at zero temperature, $\bar{\Delta}_{0,n}$, and (b) its ratio to the critical temperature, $\bar{\Delta}_{0,n} / T_c$, as functions of the baryon number density. For comparison, results for both the slab phases and uniform nuclear matter are presented in each panel.
Previous studies in this context~\cite{drissi2022} have shown that $\bar{\Delta}_{0,n}$ exhibits a parabolic dependence on density, with a maximum around $n_B = 0.01$-$0.02~\mathrm{fm}^{-3}$. 
However, $\bar{\Delta}_{0,n}$ in the present study increases monotonically with temperature in both the slab and uniform cases.
One possible reason for this discrepancy is that the density dependence of the pairing interaction is neglected in the present work. The adopted constant coupling strength, $g_0 = 200~\mathrm{MeV~fm}$, is a standard value often used in calculations for finite nuclei~\cite{yu2003,bulgac2018}, yielding pairing gaps on the order of 1 MeV. Nonetheless, more sophisticated approaches exist: for instance, one may introduce a density-dependent coupling that reproduces the expected density dependence of the pairing gap~\cite{wlazlowski2016a}, or use a modified density-dependent delta interaction, as proposed in Ref.~\cite{okihashi2021}. Incorporating such refinements may help to reconcile the discrepancy between the current results and those predicted by many-body theories.
On the other hand, the ratio $\bar{\Delta}_{0,n} / T_c$ remains nearly constant, around 1.8-1.9, across all examined baryon densities. This value is consistent with results obtained using many-body approximations within the BCS framework (see Fig.~2 in Ref.~\cite{drissi2022}).

\subsection{Magnetic field effects}

In this section, we investigate the influence of external magnetic field on the properties of nuclear matter. We will show the results at both zero and finite temperatures to examine an interplay of effects of finite temperature and magnetic field.

\subsubsection{Structural effects}
\begin{figure}
    \centering\vspace{-10mm}
    \includegraphics[width=\columnwidth]{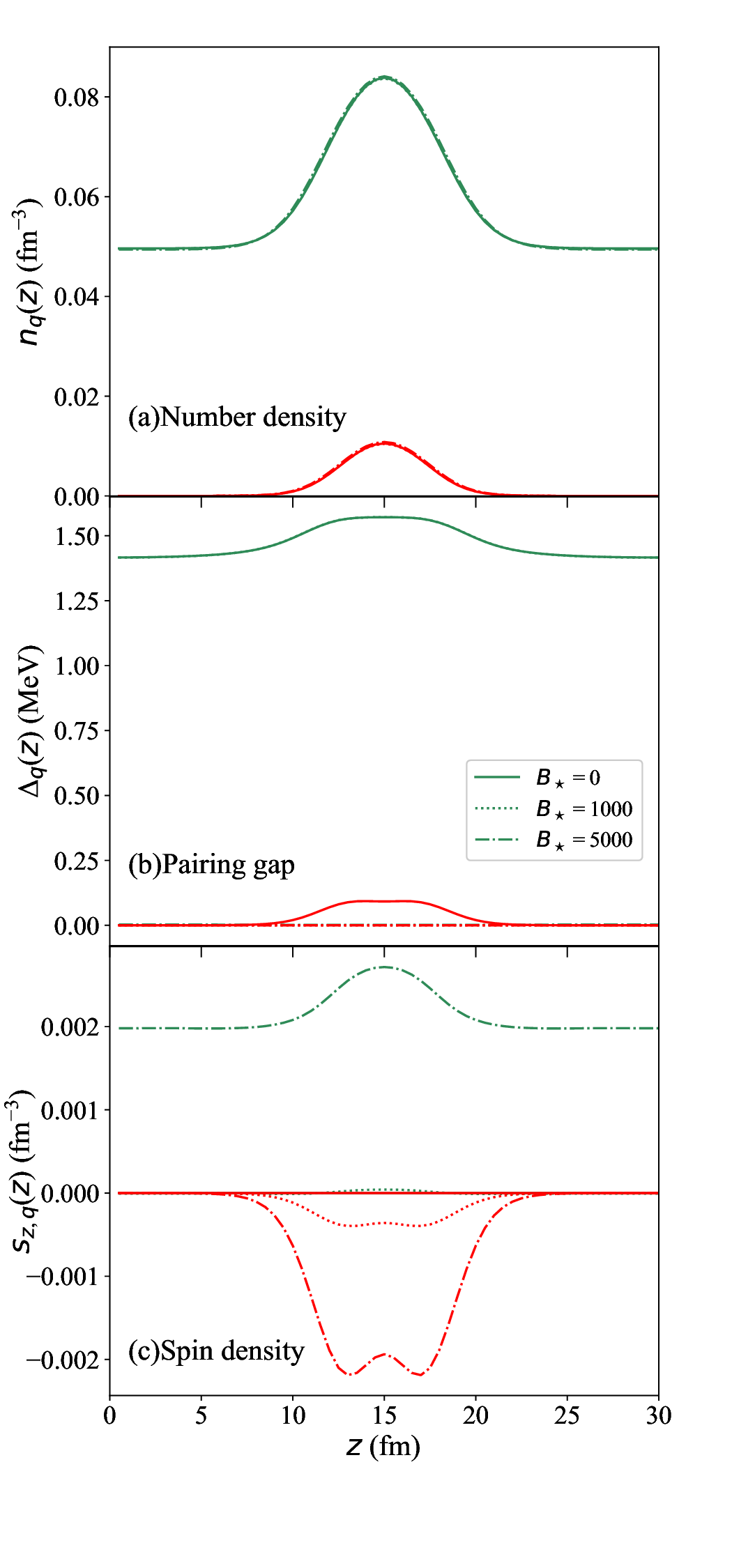}\vspace{-8mm}
    \caption{(a) The number density $n_q$ fm$^{-3}$, (b) pairing field $\Delta_q$ MeV, and (c) the $z$ component of the spin density $s_{z,q}$ fm$^{-3}$ are shown as a function of the spatial coordinate $z$ in three cases with various magnetic field strengths $B_\star = 0$, $1000$, $5000$.
    }
    \label{fig:magdens}
\end{figure}

Figure~\ref{fig:magdens} presents the nucleon densities (panel (a)) and spin densities (panel (c)) as functions of the spatial coordinate $z$ under various magnetic field strengths.
As shown in panel (a), the density profiles of the inner crust matter remain largely unchanged, even under extremely strong magnetic fields such as $B = 5000 B_\star \sim 2 \times 10^{17}~\mathrm{G}$. This suggests that structural phase transitions induced by magnetic field effects are unlikely to occur within the physically relevant range of field strengths.
In contrast, panel (c) reveals a clear evolution of the spin densities. In the absence of magnetic fields, both neutron and proton spin densities vanish. Under a moderate magnetic field of $B = 1000 B_\star$, finite spin densities emerge—although the neutron component remains small—and further increase substantially at $B = 5000 B_\star$. These findings indicate the presence of a magnetic phase transition from a paramagnetic to a ferromagnetic state, likely occurring within the magnetic field range of $B \approx 10^{15}$–$10^{16}~\mathrm{G}$.
In the following subsections, we examine the phase structure of inner crust matter under magnetic fields in greater detail, focusing on selected field strengths.

\subsubsection{Pairing phase transition}



Figure~\ref{fig:tcomp-delta-spin}(a) shows the average magnitude of the neutron pairing field, defined as $\overline{\Delta}_n = \int\Delta_n(z)n_n(z)\,\dd z/N_n$, as a function of magnetic field strength for $n_\text{B} = 0.05\,\mathrm{fm}^{-3}$ and $a = 30\,\mathrm{fm}$. Results are shown for four representative temperatures: $k_\text{B}T = 0$, $10$, $100$, and $1000$~keV, indicated by solid, dashed, dotted, and dash-dotted lines with circle, upward triangle, downward triangle, and square symbols, respectively.

As seen in the figure, the average neutron pairing gap remains nearly constant up to a critical magnetic field strength, beyond which it abruptly drops to zero. This drop indicates the breaking of Cooper pairs due to the magnetic field. Notably, in the $k_\text{B}T = 1000$\,keV case, neutron superfluidity is already absent even without a magnetic field, since the temperature exceeds the critical value $T_\text{c}^\text{sf} = 0.73$\,MeV.
Despite the seemingly abrupt nature of the transition, the gradual decrease in pairing observed at $k_\text{B}T=100$~keV suggests that it is more appropriately classified as a second-order phase transition.
Thus, the transition from the superfluid to the normal phase is triggered by the magnetic field and occurs at a temperature-dependent critical field strength, signaling the onset of spin polarization.

\subsubsection{Spin polarization}

\begin{figure}
    \centering\vspace{-10mm}
    \includegraphics[width=\columnwidth]{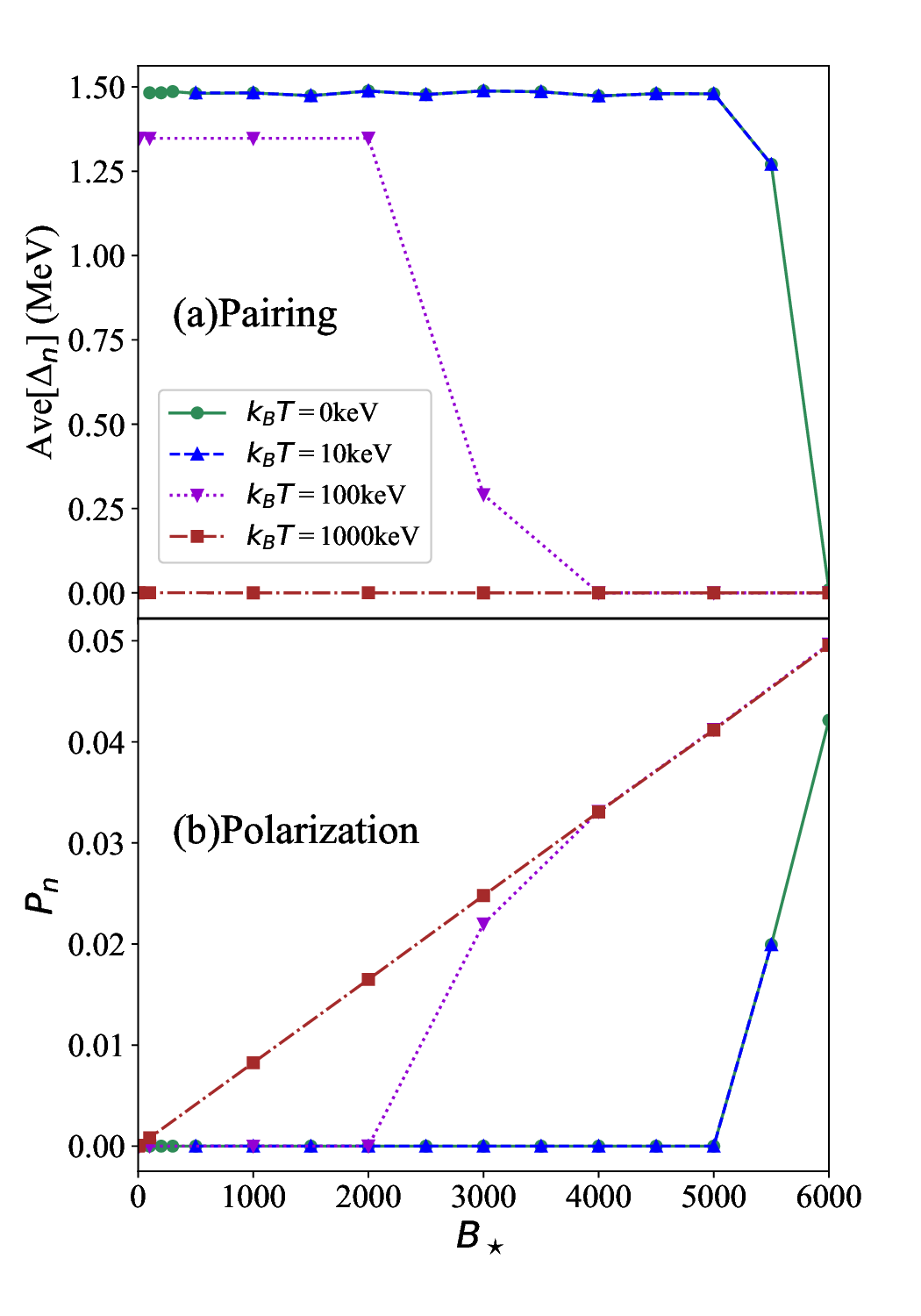}\vspace{-8mm}
    \caption{(a) Average magnitude of the neutron pairing field, $\overline{\Delta}_n$, and (b) total spin polarization, $P_n$, of neutrons are shown as a function of the magnetic field strength, $B_\star$, at four representative temperatures, $k_\text{B}T=0$, $10$, $100$, and $1000\,\mathrm{keV}$. In ascending order of temperatures, circle, upward triangle, downward triangle, and rectangular symbols connected with solid, dashed, dotted, and dash-dotted lines are used, respectively.
    }
    \label{fig:tcomp-delta-spin}
\end{figure}

In the inner crust of neutron stars, neutrons form a superfluid phase associated with $s$-wave spin-singlet ($^1\text{S}_0$) pairing. In contrast, coupling to an external magnetic field tends to align the nucleon spins parallel to the field direction, leading to spin polarization. This gives rise to a clear competition between nuclear pairing correlations and magnetic-field-induced spin alignment.
By incorporating magnetic-field effects into the finite-temperature superfluid local density approximation (SLDA) framework, we are now able to investigate this competition in a fully microscopic and self-consistent manner.

Figure~\ref{fig:tcomp-delta-spin}(b) displays the total neutron spin polarization, defined as $P_n = (N_{n,\uparrow} - N_{n,\downarrow}) / (N_{n,\uparrow} + N_{n,\downarrow})$, as a function of the magnetic field strength $B_\star$. The four curves, distinguished by line styles and symbols, correspond to four representative temperatures, in the same manner as in panel (a).
As shown in the figure, the system remains unpolarized ($P_n = 0$) up to a certain critical magnetic field strength. Notably, protons exhibit finite spin polarization even at relatively low fields (around $B_\star \approx 200$), due to their smaller pairing gap. Beyond the critical field, the neutron system exhibits a sudden onset of small but finite spin polarization. This critical point coincides with the magnetic field strengths at which a sharp drop in the average neutron pairing gap was observed in Fig.~\ref{fig:tcomp-delta-spin}(a).
After the onset of polarization, both neutrons and protons exhibit an almost linear increase in spin polarization with increasing magnetic field strength. It is also evident from the figure that the critical field for the onset of spin polarization decreases significantly as temperature increases. At $k_\text{B}T = 1$~MeV, the pairing gap has vanished even in the absence of a magnetic field, and the polarization increases linearly with field strength from the outset.
The strong correlation between the suppression of the pairing gap and the emergence of spin polarization suggests a close interplay between nucleon pairing correlations and magnetic-field-induced spin polarization.

\begin{figure}[tbp]
    \centering
    \includegraphics[width=\columnwidth]{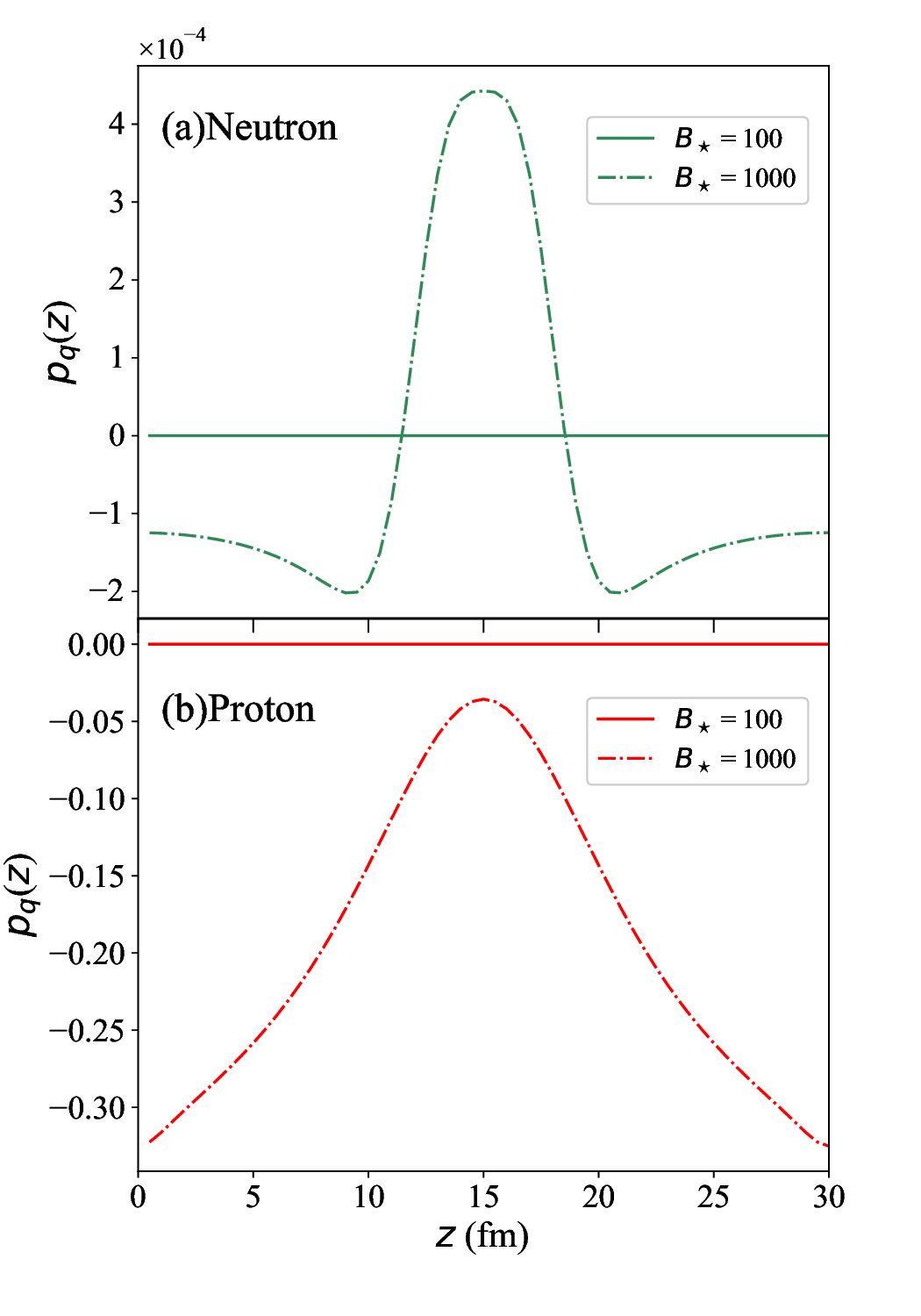}\vspace{-7mm}
    \caption{Distributions of the local spin polarization of neutrons, $p_n(z)$, and protons, $p_p(z)$, are shown in panels (a) and (b), respectively, as a function of $z$ coordinate for two magnetic field strengths, $B_\star=100$ and $1000$. Results for $B_\star=100$ are shown by solid lines, while those for $B_\star=1000$ are shown by dashed lines.
    }
    \label{fig:spindens}
\end{figure}

Let us now examine in more detail the internal structure of the system under an external magnetic field. Figures~\ref{fig:spindens}(a) and \ref{fig:spindens}(b) show the local spin polarization of neutrons and protons, respectively, as a function of the spatial coordinate $z$, for magnetic field strengths $B_\star = 100$ and $1000$ at zero temperature. As indicated in Fig.~\ref{fig:tcomp-delta-spin}, the $B_\star = 100$ case corresponds to the paramagnetic phase for both neutrons and protons, whereas at $B_\star = 1000$, only protons exhibit spin polarization.
Although protons appear to exhibit local polarization outside the nuclear slabs in the $B_\star = 1000$ case, this effect should be disregarded, as there are no free protons in that region, as confirmed in Fig.~\ref{fig:magdens}(c).
From Figs.~\ref{fig:spindens}(a) and \ref{fig:spindens}(b), it is evident that neither neutrons nor protons are polarized at any spatial point (i.e., $p_q(z) = 0$) in the $B_\star = 100$ case (solid lines). In contrast, the $B_\star = 1000$ results reveal several notable features of nuclear matter under strong magnetic fields. First, in the central region containing the nuclear slab, neutrons and protons are polarized in opposite directions ($p_n > 0$ and $p_p < 0$), reflecting the opposite signs of their $g$-factors. In the outer region containing dripped neutrons, however, both species are polarized in the same direction ($p_n < 0$ and $p_p < 0$), indicating that the local neutron polarization changes sign across the slab boundary.
It is important to note that, despite the presence of local polarization, neutrons remain unpolarized overall even at $B_\star = 1000$, as shown in Fig.~\ref{fig:tcomp-delta-spin}.

\begin{figure}
    \centering
    \includegraphics[width=\columnwidth]{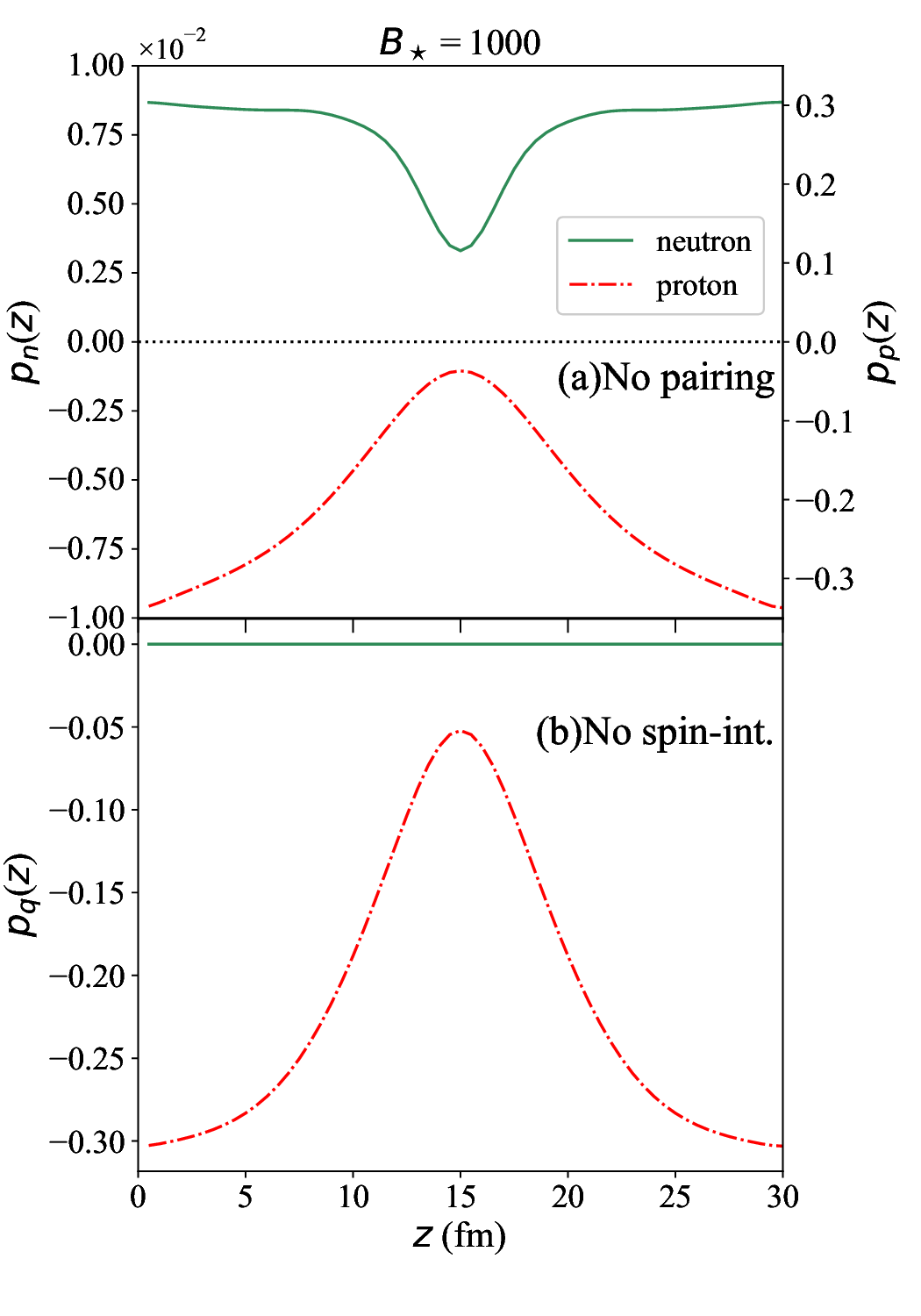}\vspace{-7mm}
    \caption{Distributions of the local spin polarization of neutrons (green solid lines) and of protons (red dash-dotted lines) are shown as a function of $z$ coordinate for the $B_\star=1000$ case. In panel (a), results without the pairing interaction are presented, while in panel (b) results without the spin-dependent interaction are shown.}
    \label{fig:nopair-spin}
\end{figure}

To investigate the origin of the observed local neutron spin polarization at $B_\star = 1000$, we performed additional calculations with two modified interaction setups. Figure~\ref{fig:nopair-spin}(a) shows the local neutron spin polarization at $B_\star = 1000$ with the pairing interaction turned off (i.e., $\Delta_q = 0$). A comparison between Figs.~\ref{fig:spindens}(a) (dash-dotted line) and \ref{fig:nopair-spin}(a) (solid line) reveals a qualitative change in the neutron spin polarization: in the absence of pairing, neutrons become spin polarized overall, and the profile of $p_n(z)$ qualitatively resembles that of $p_p(z)$, albeit with the opposite sign.
For further comparison, Fig.~\ref{fig:nopair-spin}(b) presents results obtained with the pairing interaction included but with spin-dependent terms omitted from the energy density functional (EDF). In this setup, all terms involving time-odd spin densities ($\bm{s}_q$) are excluded. In stark contrast to the previous case, neutrons remain completely unpolarized throughout the entire space ($p_n(z) = 0$), as the spin-singlet pairing correlations are fully preserved.
These results demonstrate that the emergence of local spin polarization in the presence of pairing arises from the interplay between spin-singlet pairing and spin-dependent interactions. While the former favors a globally unpolarized state, the latter tend to reduce the isoscalar spin density ($s_{z,0} = s_{z,n} + s_{z,p}$), thereby encouraging neutrons and protons to become polarized in opposite directions.
We note that the spin-dependent interactions originate from time-odd terms in the EDF, characterized by the coupling constants $C_t^{\bm{s}}$, $C_{t\text{D}}^{\bm{s}}$, and $C_t^{\bm{T}}$, which are generally less well constrained compared to their time-even counterparts. A systematic investigation into the functional dependence and physical implications of these spin-dependent terms remains an important direction for future work.

In summary, our superfluid band theory calculations indicate that, even under magnetic fields below the critical strength at which Cooper pairs are broken, neutrons located inside and outside nuclear clusters can become spin-polarized in opposite directions. Nevertheless, the system as a whole retains zero net polarization.
Notably, the persistence of local spin polarization without a suppression of the pairing gap $\Delta$ suggests that neutron Cooper pairs can form between spatially separated regions—specifically, with one neutron inside and the other outside the cluster. This phenomenon implies the possibility of long-range pairing correlations extending across the nuclear interface.
We also note that $B_\star = 1000$ corresponds to $B \sim 4.41 \times 10^{16}~\mathrm{G}$, a field strength potentially realized in magnetars. In such extreme environments, the presence of spatially varying spin polarization in the inner crust may influence various physical processes, including thermal conductivity, neutrino emission, and transport properties.

\section{Summary and Prospect}\label{Sec:Summary}

In this work, we have extended the fully microscopic superfluid band theory based on Kohn-Sham density functional theory (DFT) for superfluid systems, commonly known as the superfluid local density approximation (SLDA), to incorporate the effects of finite temperature and magnetic fields. 
For finite-temperature systems, densities are computed using a mixture of the $u$- and $v$-components of quasi-particle wave functions, in a manner analogous to the finite-temperature Hartree-Fock-Bogoliubov framework. For systems under magnetic fields, the chemical potential of relativistic electrons is evaluated by incorporating Landau-Rabi quantization, which influences the neutron-proton composition via the condition of $\beta$-equilibrium. Additionally, the magnetic interaction with nucleons is treated by introducing the coupling between the external magnetic field and the magnetic moments of the nucleons into the single-particle Hamiltonian. We note that we have neglected possible effects of Landau quantization of protons, which may play an important role under a superstrong magnetic field on the order of $10^{18}$\,G \cite{broderick2000}. This extended framework enables fully microscopic calculations of neutron star matter under realistic thermal and magnetic conditions. Its feasibility has been demonstrated through application to the slab phase of nuclear matter.

From our finite-temperature calculations, we have identified two distinct types of phase transitions: a pairing transition for neutrons occurring around $k_\text{B}T = 1$\,MeV, and a structural transition—corresponding to the melting of nuclear clusters—around $k_\text{B}T = 4$\,MeV.
These critical temperatures exhibit systematic dependence on the baryon number density. Specifically, as the density increases, the critical temperature for the pairing transition increases, while that for the structural transition decreases. The former trend indicates enhanced pairing correlations among neutrons at higher densities, whereas the latter reflects shape changes in the clusters, which become increasingly diffuse and approach uniform nuclear matter as the baryon number density rises.

From our finite magnetic-field calculations, we have revealed a rich variety of phases in the inner crust of neutron stars under external magnetic fields. Up to a certain critical field strength, neutrons maintain $^1\text{S}_0$ superfluidity, which suppresses spin polarization. This critical magnetic-field strength decreases with increasing temperature due to thermal suppression of pairing correlations, a behavior reminiscent of antiferromagnetism.
Beyond the critical field strength, the $^1\text{S}_0$ Cooper pairs are broken, and the neutron spin polarization increases nearly linearly with the magnetic field, exhibiting paramagnetic behavior. A detailed analysis of the spatial distribution of spin polarization further reveals that even below the critical field, neutrons are locally polarized in opposite directions inside and outside the nuclear clusters. Specifically, neutrons inside the cluster are polarized oppositely to protons, consistent with the opposite signs of their $g$-factors, while neutrons outside the cluster are polarized in the same direction as protons. As a result, the total neutron spin polarization in the system remains nearly zero.
This implies that neutrons inside the clusters exhibit ferromagnetic behavior, whereas dripped neutrons outside the clusters behave as if they possess diamagnetic characteristics. We have shown that the emergence of local spin polarization arises from the interplay between pairing and spin-dependent interactions. While the onset of polarization appears abrupt—resembling a first-order phase transition—at zero temperature, it becomes smoother at finite temperatures.
These findings may have significant implications for the physics of neutron stars under realistic astrophysical conditions, where the effects of finite temperature and strong magnetic fields, such as those present during the cooling of proto-neutron stars~\cite{sumiyoshi2023}, are non-negligible.

As a future work, it would be interesting to investigate how the structure and properties of the system change when the magnetic field direction is changed. 
By applying magnetic fields in various directions and analyzing the resulting equation of state, it may become possible to determine the preferred orientation of slab sequences under strong magnetic fields. This information is expected to be particularly important for understanding neutrino opacity as well as orientations of pasta structures.
Additionally, we indeed plan to extend our theoretical framework to two- and three-dimensional geometries. Such an extension will allow for a more comprehensive and self-consistent description of neutron star matter that incorporates the effects of finite temperature, magnetic fields, superfluidity, and band structure on the same footing.
In the outer core of neutron stars, neutrons are expected to form a spin-triplet $p$-wave ($^3\text{P}_2$) superfluid, which may be more robust against external magnetic fields. This, in turn, could significantly influence the behavior of spin polarization. Our formalism can potentially be extended to include $^3\text{P}_2$ pairing by incorporating spin-current pair densities, as discussed in Ref.~\cite{hinohara2024}. While such spin-triplet correlations cannot arise in one-dimensional systems due to the absence of spin–orbit coupling, they are expected to play a pivotal role in higher-dimensional geometries.
Furthermore, insights from cold-atom physics suggest that spin polarization can give rise to exotic phases such as the Larkin–Ovchinnikov–Fulde–Ferrell (LOFF) phase~\cite{fulde1964, larkin1964, casalbuoni2004}, as well as spin-polarized droplets known as \textit{ferrons}~\cite{magierski2019, magierski2021} and their complex spatial patterns~\cite{tuzemen2023}. In nuclear systems, it would also be intriguing to explore possible topological structures, such as Skyrmion crystals~\cite{hayami2021, lee2022}, which may emerge under strong spin polarization.
By taking into account these various internal and external effects, we aim to establish a comprehensive microscopic framework for investigating the properties of neutron star matter under realistic astrophysical conditions, spanning from their birth to their later evolutionary stages.


\section*{Acknowledgments}
We would like to thank Kenichi Yoshida (RCNP, Osaka University) and Takashi Nakatsukasa (University of Tsukuba) for valuable discussions. One of the authors (K.Y.) appreciate the support from the Hiki Foundation, Institute of Science Tokyo. This work is supported by the JSPS Research Fellow, Grant No.~24KJ1110, as well as JSPS Grant-in-Aid for Scientific Research, Grants No.~23K03410 and No.~23K25864. This work mainly used computational resources of the Yukawa-21 supercomputer at Yukawa Institute for Theoretical Physics (YITP), Kyoto University. This work also used (in part) the computational resources of TSUBAME4.0 at Institute of Science Tokyo, through the HPCI System Project, Project ID: hp230180, hp240183, and hp250097.

\bibliographystyle{apsrev4-2}
\bibliography{ftband}

\begin{thebibliography}{94}%
\makeatletter
\providecommand \@ifxundefined [1]{%
 \@ifx{#1\undefined}
}%
\providecommand \@ifnum [1]{%
 \ifnum #1\expandafter \@firstoftwo
 \else \expandafter \@secondoftwo
 \fi
}%
\providecommand \@ifx [1]{%
 \ifx #1\expandafter \@firstoftwo
 \else \expandafter \@secondoftwo
 \fi
}%
\providecommand \natexlab [1]{#1}%
\providecommand \enquote  [1]{``#1''}%
\providecommand \bibnamefont  [1]{#1}%
\providecommand \bibfnamefont [1]{#1}%
\providecommand \citenamefont [1]{#1}%
\providecommand \href@noop [0]{\@secondoftwo}%
\providecommand \href [0]{\begingroup \@sanitize@url \@href}%
\providecommand \@href[1]{\@@startlink{#1}\@@href}%
\providecommand \@@href[1]{\endgroup#1\@@endlink}%
\providecommand \@sanitize@url [0]{\catcode `\\12\catcode `\$12\catcode `\&12\catcode `\#12\catcode `\^12\catcode `\_12\catcode `\%12\relax}%
\providecommand \@@startlink[1]{}%
\providecommand \@@endlink[0]{}%
\providecommand \url  [0]{\begingroup\@sanitize@url \@url }%
\providecommand \@url [1]{\endgroup\@href {#1}{\urlprefix }}%
\providecommand \urlprefix  [0]{URL }%
\providecommand \Eprint [0]{\href }%
\providecommand \doibase [0]{https://doi.org/}%
\providecommand \selectlanguage [0]{\@gobble}%
\providecommand \bibinfo  [0]{\@secondoftwo}%
\providecommand \bibfield  [0]{\@secondoftwo}%
\providecommand \translation [1]{[#1]}%
\providecommand \BibitemOpen [0]{}%
\providecommand \bibitemStop [0]{}%
\providecommand \bibitemNoStop [0]{.\EOS\space}%
\providecommand \EOS [0]{\spacefactor3000\relax}%
\providecommand \BibitemShut  [1]{\csname bibitem#1\endcsname}%
\let\auto@bib@innerbib\@empty
\bibitem [{\citenamefont {Nakatsukasa}\ \emph {et~al.}(2016)\citenamefont {Nakatsukasa}, \citenamefont {Matsuyanagi}, \citenamefont {Matsuo},\ and\ \citenamefont {Yabana}}]{nakatsukasa2016a}%
  \BibitemOpen
  \bibfield  {author} {\bibinfo {author} {\bibfnamefont {T.}~\bibnamefont {Nakatsukasa}}, \bibinfo {author} {\bibfnamefont {K.}~\bibnamefont {Matsuyanagi}}, \bibinfo {author} {\bibfnamefont {M.}~\bibnamefont {Matsuo}},\ and\ \bibinfo {author} {\bibfnamefont {K.}~\bibnamefont {Yabana}},\ }\href {https://link.aps.org/doi/10.1103/RevModPhys.88.045004} {\bibfield  {journal} {\bibinfo  {journal} {Rev. Mod. Phys.}\ }\textbf {\bibinfo {volume} {88}},\ \bibinfo {pages} {045004} (\bibinfo {year} {2016})}\BibitemShut {NoStop}%
\bibitem [{\citenamefont {Col^^c3^^b2}(2020)}]{Col^^c3^^b22020}%
  \BibitemOpen
  \bibfield  {author} {\bibinfo {author} {\bibfnamefont {G.}~\bibnamefont {Col^^c3^^b2}},\ }\href {https://doi.org/10.1080/23746149.2020.1740061} {\bibfield  {journal} {\bibinfo  {journal} {Advances in Physics: X}\ }\textbf {\bibinfo {volume} {5}},\ \bibinfo {pages} {1740061} (\bibinfo {year} {2020})}\BibitemShut {NoStop}%
\bibitem [{\citenamefont {Bender}\ \emph {et~al.}(2003)\citenamefont {Bender}, \citenamefont {Heenen},\ and\ \citenamefont {Reinhard}}]{bender2003}%
  \BibitemOpen
  \bibfield  {author} {\bibinfo {author} {\bibfnamefont {M.}~\bibnamefont {Bender}}, \bibinfo {author} {\bibfnamefont {P.-H.}\ \bibnamefont {Heenen}},\ and\ \bibinfo {author} {\bibfnamefont {P.-G.}\ \bibnamefont {Reinhard}},\ }\href {https://link.aps.org/doi/10.1103/RevModPhys.75.121} {\bibfield  {journal} {\bibinfo  {journal} {Rev. Mod. Phys.}\ }\textbf {\bibinfo {volume} {75}},\ \bibinfo {pages} {121} (\bibinfo {year} {2003})}\BibitemShut {NoStop}%
\bibitem [{\citenamefont {Ravenhall}\ \emph {et~al.}(1983)\citenamefont {Ravenhall}, \citenamefont {Pethick},\ and\ \citenamefont {Wilson}}]{Ravenhall1983}%
  \BibitemOpen
  \bibfield  {author} {\bibinfo {author} {\bibfnamefont {D.~G.}\ \bibnamefont {Ravenhall}}, \bibinfo {author} {\bibfnamefont {C.~J.}\ \bibnamefont {Pethick}},\ and\ \bibinfo {author} {\bibfnamefont {J.~R.}\ \bibnamefont {Wilson}},\ }\href {https://doi.org/10.1103/PhysRevLett.50.2066} {\bibfield  {journal} {\bibinfo  {journal} {Phys. Rev. Lett.}\ }\textbf {\bibinfo {volume} {50}},\ \bibinfo {pages} {2066} (\bibinfo {year} {1983})}\BibitemShut {NoStop}%
\bibitem [{\citenamefont {Hashimoto}\ \emph {et~al.}(1984)\citenamefont {Hashimoto}, \citenamefont {Seki},\ and\ \citenamefont {Yamada}}]{Hashimoto1984}%
  \BibitemOpen
  \bibfield  {author} {\bibinfo {author} {\bibfnamefont {M.}~\bibnamefont {Hashimoto}}, \bibinfo {author} {\bibfnamefont {H.}~\bibnamefont {Seki}},\ and\ \bibinfo {author} {\bibfnamefont {M.}~\bibnamefont {Yamada}},\ }\href {https://doi.org/10.1143/PTP.71.320} {\bibfield  {journal} {\bibinfo  {journal} {Progress of Theoretical Physics}\ }\textbf {\bibinfo {volume} {71}},\ \bibinfo {pages} {320} (\bibinfo {year} {1984})}\BibitemShut {NoStop}%
\bibitem [{\citenamefont {Magierski}\ and\ \citenamefont {Heenen}(2002)}]{magierski2002}%
  \BibitemOpen
  \bibfield  {author} {\bibinfo {author} {\bibfnamefont {P.}~\bibnamefont {Magierski}}\ and\ \bibinfo {author} {\bibfnamefont {P.-H.}\ \bibnamefont {Heenen}},\ }\href {https://link.aps.org/doi/10.1103/PhysRevC.65.045804} {\bibfield  {journal} {\bibinfo  {journal} {Phys. Rev. C}\ }\textbf {\bibinfo {volume} {65}},\ \bibinfo {pages} {045804} (\bibinfo {year} {2002})}\BibitemShut {NoStop}%
\bibitem [{\citenamefont {G{\"o}gelein}\ \emph {et~al.}(2008)\citenamefont {G{\"o}gelein}, \citenamefont {van Dalen}, \citenamefont {Fuchs},\ and\ \citenamefont {M{\"u}ther}}]{gogelein2008}%
  \BibitemOpen
  \bibfield  {author} {\bibinfo {author} {\bibfnamefont {P.}~\bibnamefont {G{\"o}gelein}}, \bibinfo {author} {\bibfnamefont {E.~N.~E.}\ \bibnamefont {van Dalen}}, \bibinfo {author} {\bibfnamefont {C.}~\bibnamefont {Fuchs}},\ and\ \bibinfo {author} {\bibfnamefont {H.}~\bibnamefont {M{\"u}ther}},\ }\href {https://link.aps.org/doi/10.1103/PhysRevC.77.025802} {\bibfield  {journal} {\bibinfo  {journal} {Phys. Rev. C}\ }\textbf {\bibinfo {volume} {77}},\ \bibinfo {pages} {025802} (\bibinfo {year} {2008})}\BibitemShut {NoStop}%
\bibitem [{\citenamefont {Newton}\ and\ \citenamefont {Stone}(2009)}]{newton2009}%
  \BibitemOpen
  \bibfield  {author} {\bibinfo {author} {\bibfnamefont {W.~G.}\ \bibnamefont {Newton}}\ and\ \bibinfo {author} {\bibfnamefont {J.~R.}\ \bibnamefont {Stone}},\ }\href {https://link.aps.org/doi/10.1103/PhysRevC.79.055801} {\bibfield  {journal} {\bibinfo  {journal} {Phys. Rev. C}\ }\textbf {\bibinfo {volume} {79}},\ \bibinfo {pages} {055801} (\bibinfo {year} {2009})}\BibitemShut {NoStop}%
\bibitem [{\citenamefont {Pais}\ and\ \citenamefont {Stone}(2012)}]{pais2012}%
  \BibitemOpen
  \bibfield  {author} {\bibinfo {author} {\bibfnamefont {H.}~\bibnamefont {Pais}}\ and\ \bibinfo {author} {\bibfnamefont {J.~R.}\ \bibnamefont {Stone}},\ }\href {https://link.aps.org/doi/10.1103/PhysRevLett.109.151101} {\bibfield  {journal} {\bibinfo  {journal} {Phys. Rev. Lett.}\ }\textbf {\bibinfo {volume} {109}},\ \bibinfo {pages} {151101} (\bibinfo {year} {2012})}\BibitemShut {NoStop}%
\bibitem [{\citenamefont {Grill}\ \emph {et~al.}(2014)\citenamefont {Grill}, \citenamefont {Pais}, \citenamefont {Provid{\^e}ncia}, \citenamefont {Vida{\~n}a},\ and\ \citenamefont {Avancini}}]{grill2014}%
  \BibitemOpen
  \bibfield  {author} {\bibinfo {author} {\bibfnamefont {F.}~\bibnamefont {Grill}}, \bibinfo {author} {\bibfnamefont {H.}~\bibnamefont {Pais}}, \bibinfo {author} {\bibfnamefont {C.}~\bibnamefont {Provid{\^e}ncia}}, \bibinfo {author} {\bibfnamefont {I.}~\bibnamefont {Vida{\~n}a}},\ and\ \bibinfo {author} {\bibfnamefont {S.~S.}\ \bibnamefont {Avancini}},\ }\href {https://link.aps.org/doi/10.1103/PhysRevC.90.045803} {\bibfield  {journal} {\bibinfo  {journal} {Phys. Rev. C}\ }\textbf {\bibinfo {volume} {90}},\ \bibinfo {pages} {045803} (\bibinfo {year} {2014})}\BibitemShut {NoStop}%
\bibitem [{\citenamefont {Schuetrumpf}\ and\ \citenamefont {Nazarewicz}(2015)}]{schuetrumpf2015b}%
  \BibitemOpen
  \bibfield  {author} {\bibinfo {author} {\bibfnamefont {B.}~\bibnamefont {Schuetrumpf}}\ and\ \bibinfo {author} {\bibfnamefont {W.}~\bibnamefont {Nazarewicz}},\ }\href {https://doi.org/10.1103/PhysRevC.92.045806} {\bibfield  {journal} {\bibinfo  {journal} {Phys. Rev. C}\ }\textbf {\bibinfo {volume} {92}},\ \bibinfo {pages} {045806} (\bibinfo {year} {2015})}\BibitemShut {NoStop}%
\bibitem [{\citenamefont {Fattoyev}\ \emph {et~al.}(2017)\citenamefont {Fattoyev}, \citenamefont {Horowitz},\ and\ \citenamefont {Schuetrumpf}}]{fattoyev2017}%
  \BibitemOpen
  \bibfield  {author} {\bibinfo {author} {\bibfnamefont {F.~J.}\ \bibnamefont {Fattoyev}}, \bibinfo {author} {\bibfnamefont {C.~J.}\ \bibnamefont {Horowitz}},\ and\ \bibinfo {author} {\bibfnamefont {B.}~\bibnamefont {Schuetrumpf}},\ }\href {https://doi.org/10.1103/PhysRevC.95.055804} {\bibfield  {journal} {\bibinfo  {journal} {Phys. Rev. C}\ }\textbf {\bibinfo {volume} {95}},\ \bibinfo {pages} {055804} (\bibinfo {year} {2017})}\BibitemShut {NoStop}%
\bibitem [{\citenamefont {Schuetrumpf}\ \emph {et~al.}(2019)\citenamefont {Schuetrumpf}, \citenamefont {Mart\'{\i}nez-Pinedo}, \citenamefont {Afibuzzaman},\ and\ \citenamefont {Aktulga}}]{schuetrumpf2019}%
  \BibitemOpen
  \bibfield  {author} {\bibinfo {author} {\bibfnamefont {B.}~\bibnamefont {Schuetrumpf}}, \bibinfo {author} {\bibfnamefont {G.}~\bibnamefont {Mart\'{\i}nez-Pinedo}}, \bibinfo {author} {\bibfnamefont {M.}~\bibnamefont {Afibuzzaman}},\ and\ \bibinfo {author} {\bibfnamefont {H.~M.}\ \bibnamefont {Aktulga}},\ }\href {https://doi.org/10.1103/PhysRevC.100.045806} {\bibfield  {journal} {\bibinfo  {journal} {Phys. Rev. C}\ }\textbf {\bibinfo {volume} {100}},\ \bibinfo {pages} {045806} (\bibinfo {year} {2019})}\BibitemShut {NoStop}%
\bibitem [{\citenamefont {Schuetrumpf}\ \emph {et~al.}(2020)\citenamefont {Schuetrumpf}, \citenamefont {{Mart{\'i}nez-Pinedo}},\ and\ \citenamefont {Reinhard}}]{schuetrumpf2020}%
  \BibitemOpen
  \bibfield  {author} {\bibinfo {author} {\bibfnamefont {B.}~\bibnamefont {Schuetrumpf}}, \bibinfo {author} {\bibfnamefont {G.}~\bibnamefont {{Mart{\'i}nez-Pinedo}}},\ and\ \bibinfo {author} {\bibfnamefont {P.-G.}\ \bibnamefont {Reinhard}},\ }\href {https://link.aps.org/doi/10.1103/PhysRevC.101.055804} {\bibfield  {journal} {\bibinfo  {journal} {Phys. Rev. C}\ }\textbf {\bibinfo {volume} {101}},\ \bibinfo {pages} {055804} (\bibinfo {year} {2020})}\BibitemShut {NoStop}%
\bibitem [{\citenamefont {Schuetrumpf}\ \emph {et~al.}(2013)\citenamefont {Schuetrumpf}, \citenamefont {Klatt}, \citenamefont {Iida}, \citenamefont {Maruhn}, \citenamefont {Mecke},\ and\ \citenamefont {Reinhard}}]{schuetrumpf2013}%
  \BibitemOpen
  \bibfield  {author} {\bibinfo {author} {\bibfnamefont {B.}~\bibnamefont {Schuetrumpf}}, \bibinfo {author} {\bibfnamefont {M.~A.}\ \bibnamefont {Klatt}}, \bibinfo {author} {\bibfnamefont {K.}~\bibnamefont {Iida}}, \bibinfo {author} {\bibfnamefont {J.~A.}\ \bibnamefont {Maruhn}}, \bibinfo {author} {\bibfnamefont {K.}~\bibnamefont {Mecke}},\ and\ \bibinfo {author} {\bibfnamefont {P.-G.}\ \bibnamefont {Reinhard}},\ }\href {https://doi.org/10.1103/PhysRevC.87.055805} {\bibfield  {journal} {\bibinfo  {journal} {Phys. Rev. C}\ }\textbf {\bibinfo {volume} {87}},\ \bibinfo {pages} {055805} (\bibinfo {year} {2013})}\BibitemShut {NoStop}%
\bibitem [{\citenamefont {Schuetrumpf}\ \emph {et~al.}(2014)\citenamefont {Schuetrumpf}, \citenamefont {Iida}, \citenamefont {Maruhn},\ and\ \citenamefont {Reinhard}}]{schuetrumpf2014}%
  \BibitemOpen
  \bibfield  {author} {\bibinfo {author} {\bibfnamefont {B.}~\bibnamefont {Schuetrumpf}}, \bibinfo {author} {\bibfnamefont {K.}~\bibnamefont {Iida}}, \bibinfo {author} {\bibfnamefont {J.~A.}\ \bibnamefont {Maruhn}},\ and\ \bibinfo {author} {\bibfnamefont {P.-G.}\ \bibnamefont {Reinhard}},\ }\href {https://doi.org/10.1103/PhysRevC.90.055802} {\bibfield  {journal} {\bibinfo  {journal} {Phys. Rev. C}\ }\textbf {\bibinfo {volume} {90}},\ \bibinfo {pages} {055802} (\bibinfo {year} {2014})}\BibitemShut {NoStop}%
\bibitem [{\citenamefont {Schuetrumpf}\ \emph {et~al.}(2015)\citenamefont {Schuetrumpf}, \citenamefont {Klatt}, \citenamefont {Iida}, \citenamefont {Schr\"oder-Turk}, \citenamefont {Maruhn}, \citenamefont {Mecke},\ and\ \citenamefont {Reinhard}}]{schuetrumpf2015a}%
  \BibitemOpen
  \bibfield  {author} {\bibinfo {author} {\bibfnamefont {B.}~\bibnamefont {Schuetrumpf}}, \bibinfo {author} {\bibfnamefont {M.~A.}\ \bibnamefont {Klatt}}, \bibinfo {author} {\bibfnamefont {K.}~\bibnamefont {Iida}}, \bibinfo {author} {\bibfnamefont {G.~E.}\ \bibnamefont {Schr\"oder-Turk}}, \bibinfo {author} {\bibfnamefont {J.~A.}\ \bibnamefont {Maruhn}}, \bibinfo {author} {\bibfnamefont {K.}~\bibnamefont {Mecke}},\ and\ \bibinfo {author} {\bibfnamefont {P.-G.}\ \bibnamefont {Reinhard}},\ }\href {https://doi.org/10.1103/PhysRevC.91.025801} {\bibfield  {journal} {\bibinfo  {journal} {Phys. Rev. C}\ }\textbf {\bibinfo {volume} {91}},\ \bibinfo {pages} {025801} (\bibinfo {year} {2015})}\BibitemShut {NoStop}%
\bibitem [{\citenamefont {Kashiwaba}\ and\ \citenamefont {Nakatsukasa}(2019)}]{kashiwaba2019}%
  \BibitemOpen
  \bibfield  {author} {\bibinfo {author} {\bibfnamefont {Y.}~\bibnamefont {Kashiwaba}}\ and\ \bibinfo {author} {\bibfnamefont {T.}~\bibnamefont {Nakatsukasa}},\ }\href {https://link.aps.org/doi/10.1103/PhysRevC.100.035804} {\bibfield  {journal} {\bibinfo  {journal} {Phys. Rev. C}\ }\textbf {\bibinfo {volume} {100}},\ \bibinfo {pages} {035804} (\bibinfo {year} {2019})}\BibitemShut {NoStop}%
\bibitem [{\citenamefont {Sekizawa}\ \emph {et~al.}(2022)\citenamefont {Sekizawa}, \citenamefont {Kobayashi},\ and\ \citenamefont {Matsuo}}]{sekizawa2022}%
  \BibitemOpen
  \bibfield  {author} {\bibinfo {author} {\bibfnamefont {K.}~\bibnamefont {Sekizawa}}, \bibinfo {author} {\bibfnamefont {S.}~\bibnamefont {Kobayashi}},\ and\ \bibinfo {author} {\bibfnamefont {M.}~\bibnamefont {Matsuo}},\ }\href {https://link.aps.org/doi/10.1103/PhysRevC.105.045807} {\bibfield  {journal} {\bibinfo  {journal} {Phys. Rev. C}\ }\textbf {\bibinfo {volume} {105}},\ \bibinfo {pages} {045807} (\bibinfo {year} {2022})}\BibitemShut {NoStop}%
\bibitem [{\citenamefont {Yoshimura}\ and\ \citenamefont {Sekizawa}(2024)}]{yoshimura2024}%
  \BibitemOpen
  \bibfield  {author} {\bibinfo {author} {\bibfnamefont {K.}~\bibnamefont {Yoshimura}}\ and\ \bibinfo {author} {\bibfnamefont {K.}~\bibnamefont {Sekizawa}},\ }\href {https://link.aps.org/doi/10.1103/PhysRevC.109.065804} {\bibfield  {journal} {\bibinfo  {journal} {Phys. Rev. C}\ }\textbf {\bibinfo {volume} {109}},\ \bibinfo {pages} {065804} (\bibinfo {year} {2024})}\BibitemShut {NoStop}%
\bibitem [{\citenamefont {Almirante}\ and\ \citenamefont {Urban}(2024{\natexlab{a}})}]{almirante2024}%
  \BibitemOpen
  \bibfield  {author} {\bibinfo {author} {\bibfnamefont {G.}~\bibnamefont {Almirante}}\ and\ \bibinfo {author} {\bibfnamefont {M.}~\bibnamefont {Urban}},\ }\href {https://link.aps.org/doi/10.1103/PhysRevC.109.045805} {\bibfield  {journal} {\bibinfo  {journal} {Phys. Rev. C}\ }\textbf {\bibinfo {volume} {109}},\ \bibinfo {pages} {045805} (\bibinfo {year} {2024}{\natexlab{a}})}\BibitemShut {NoStop}%
\bibitem [{\citenamefont {Almirante}\ and\ \citenamefont {Urban}(2024{\natexlab{b}})}]{almirante2024a}%
  \BibitemOpen
  \bibfield  {author} {\bibinfo {author} {\bibfnamefont {G.}~\bibnamefont {Almirante}}\ and\ \bibinfo {author} {\bibfnamefont {M.}~\bibnamefont {Urban}},\ }\href {https://link.aps.org/doi/10.1103/PhysRevC.110.065802} {\bibfield  {journal} {\bibinfo  {journal} {Phys. Rev. C}\ }\textbf {\bibinfo {volume} {110}},\ \bibinfo {pages} {065802} (\bibinfo {year} {2024}{\natexlab{b}})}\BibitemShut {NoStop}%
\bibitem [{\citenamefont {Ashcroft}\ and\ \citenamefont {Mermin}(1976)}]{ashcroft1976}%
  \BibitemOpen
  \bibfield  {author} {\bibinfo {author} {\bibfnamefont {N.~W.}\ \bibnamefont {Ashcroft}}\ and\ \bibinfo {author} {\bibfnamefont {N.~D.}\ \bibnamefont {Mermin}},\ }\href {https://cds.cern.ch/record/102652} {\emph {\bibinfo {title} {{Solid state physics}}}}\ (\bibinfo  {publisher} {Holt, Rinehart and Winston},\ \bibinfo {address} {New York, NY},\ \bibinfo {year} {1976})\BibitemShut {NoStop}%
\bibitem [{\citenamefont {Carter}\ \emph {et~al.}(2005)\citenamefont {Carter}, \citenamefont {Chamel},\ and\ \citenamefont {Haensel}}]{carter2005}%
  \BibitemOpen
  \bibfield  {author} {\bibinfo {author} {\bibfnamefont {B.}~\bibnamefont {Carter}}, \bibinfo {author} {\bibfnamefont {N.}~\bibnamefont {Chamel}},\ and\ \bibinfo {author} {\bibfnamefont {P.}~\bibnamefont {Haensel}},\ }\href {https://www.sciencedirect.com/science/article/pii/S0375947404011790} {\bibfield  {journal} {\bibinfo  {journal} {Nuclear Physics A}\ }\textbf {\bibinfo {volume} {748}},\ \bibinfo {pages} {675} (\bibinfo {year} {2005})}\BibitemShut {NoStop}%
\bibitem [{\citenamefont {Chamel}(2005)}]{chamel2005}%
  \BibitemOpen
  \bibfield  {author} {\bibinfo {author} {\bibfnamefont {N.}~\bibnamefont {Chamel}},\ }\href {https://www.sciencedirect.com/science/article/pii/S0375947404009200} {\bibfield  {journal} {\bibinfo  {journal} {Nuclear Physics A}\ }\textbf {\bibinfo {volume} {747}},\ \bibinfo {pages} {109} (\bibinfo {year} {2005})}\BibitemShut {NoStop}%
\bibitem [{\citenamefont {Chamel}(2012)}]{chamel2012}%
  \BibitemOpen
  \bibfield  {author} {\bibinfo {author} {\bibfnamefont {N.}~\bibnamefont {Chamel}},\ }\href {https://link.aps.org/doi/10.1103/PhysRevC.85.035801} {\bibfield  {journal} {\bibinfo  {journal} {Phys. Rev. C}\ }\textbf {\bibinfo {volume} {85}},\ \bibinfo {pages} {035801} (\bibinfo {year} {2012})}\BibitemShut {NoStop}%
\bibitem [{\citenamefont {Chamel}(2017)}]{chamel2017}%
  \BibitemOpen
  \bibfield  {author} {\bibinfo {author} {\bibfnamefont {N.}~\bibnamefont {Chamel}},\ }\href {https://doi.org/10.1007/s10909-017-1815-x} {\bibfield  {journal} {\bibinfo  {journal} {J Low Temp Phys}\ }\textbf {\bibinfo {volume} {189}},\ \bibinfo {pages} {328} (\bibinfo {year} {2017})}\BibitemShut {NoStop}%
\bibitem [{\citenamefont {Andersson}\ \emph {et~al.}(2012)\citenamefont {Andersson}, \citenamefont {Glampedakis}, \citenamefont {Ho},\ and\ \citenamefont {Espinoza}}]{andersson2012a}%
  \BibitemOpen
  \bibfield  {author} {\bibinfo {author} {\bibfnamefont {N.}~\bibnamefont {Andersson}}, \bibinfo {author} {\bibfnamefont {K.}~\bibnamefont {Glampedakis}}, \bibinfo {author} {\bibfnamefont {W.~C.~G.}\ \bibnamefont {Ho}},\ and\ \bibinfo {author} {\bibfnamefont {C.~M.}\ \bibnamefont {Espinoza}},\ }\href {https://link.aps.org/doi/10.1103/PhysRevLett.109.241103} {\bibfield  {journal} {\bibinfo  {journal} {Phys. Rev. Lett.}\ }\textbf {\bibinfo {volume} {109}},\ \bibinfo {pages} {241103} (\bibinfo {year} {2012})}\BibitemShut {NoStop}%
\bibitem [{\citenamefont {Chamel}(2013)}]{chamel2013}%
  \BibitemOpen
  \bibfield  {author} {\bibinfo {author} {\bibfnamefont {N.}~\bibnamefont {Chamel}},\ }\href {https://link.aps.org/doi/10.1103/PhysRevLett.110.011101} {\bibfield  {journal} {\bibinfo  {journal} {Phys. Rev. Lett.}\ }\textbf {\bibinfo {volume} {110}},\ \bibinfo {pages} {011101} (\bibinfo {year} {2013})}\BibitemShut {NoStop}%
\bibitem [{\citenamefont {Haskell}\ and\ \citenamefont {Melatos}(2015)}]{haskell2015a}%
  \BibitemOpen
  \bibfield  {author} {\bibinfo {author} {\bibfnamefont {B.}~\bibnamefont {Haskell}}\ and\ \bibinfo {author} {\bibfnamefont {A.}~\bibnamefont {Melatos}},\ }\href {https://www.worldscientific.com/doi/abs/10.1142/S0218271815300086} {\bibfield  {journal} {\bibinfo  {journal} {Int. J. Mod. Phys. D}\ }\textbf {\bibinfo {volume} {24}},\ \bibinfo {pages} {1530008} (\bibinfo {year} {2015})}\BibitemShut {NoStop}%
\bibitem [{\citenamefont {Watanabe}\ and\ \citenamefont {Pethick}(2017)}]{watanabe2017}%
  \BibitemOpen
  \bibfield  {author} {\bibinfo {author} {\bibfnamefont {G.}~\bibnamefont {Watanabe}}\ and\ \bibinfo {author} {\bibfnamefont {C.~J.}\ \bibnamefont {Pethick}},\ }\href {https://link.aps.org/doi/10.1103/PhysRevLett.119.062701} {\bibfield  {journal} {\bibinfo  {journal} {Phys. Rev. Lett.}\ }\textbf {\bibinfo {volume} {119}},\ \bibinfo {pages} {062701} (\bibinfo {year} {2017})}\BibitemShut {NoStop}%
\bibitem [{\citenamefont {Minami}\ and\ \citenamefont {Watanabe}(2022)}]{Watanabe2022}%
  \BibitemOpen
  \bibfield  {author} {\bibinfo {author} {\bibfnamefont {Y.}~\bibnamefont {Minami}}\ and\ \bibinfo {author} {\bibfnamefont {G.}~\bibnamefont {Watanabe}},\ }\href {https://doi.org/10.1103/PhysRevResearch.4.033141} {\bibfield  {journal} {\bibinfo  {journal} {Phys. Rev. Res.}\ }\textbf {\bibinfo {volume} {4}},\ \bibinfo {pages} {033141} (\bibinfo {year} {2022})}\BibitemShut {NoStop}%
\bibitem [{\citenamefont {Almirante}\ and\ \citenamefont {Urban}(2025)}]{almirante2025}%
  \BibitemOpen
  \bibfield  {author} {\bibinfo {author} {\bibfnamefont {G.}~\bibnamefont {Almirante}}\ and\ \bibinfo {author} {\bibfnamefont {M.}~\bibnamefont {Urban}},\ }\href {https://arxiv.org/abs/2503.21635} {\bibinfo {title} {Superfluid density in linear response theory : pulsar glitches from the inner crust of neutron stars}} (\bibinfo {year} {2025}),\ \Eprint {https://arxiv.org/abs/2503.21635} {arXiv:2503.21635 [nucl-th]} \BibitemShut {NoStop}%
\bibitem [{\citenamefont {Chamel}(2025)}]{Chamel2025}%
  \BibitemOpen
  \bibfield  {author} {\bibinfo {author} {\bibfnamefont {N.}~\bibnamefont {Chamel}},\ }\href {https://arxiv.org/abs/2412.05599} {\bibinfo {title} {Superfluid fraction in the crystalline crust of a neutron star: role of bcs pairing}} (\bibinfo {year} {2025}),\ \Eprint {https://arxiv.org/abs/2412.05599} {arXiv:2412.05599 [astro-ph.HE]} \BibitemShut {NoStop}%
\bibitem [{\citenamefont {{Lassaut}}\ \emph {et~al.}(1987)\citenamefont {{Lassaut}}, \citenamefont {{Flocard}}, \citenamefont {{Bonche}}, \citenamefont {{Heenen}},\ and\ \citenamefont {{Suraud}}}]{Lassaut1987}%
  \BibitemOpen
  \bibfield  {author} {\bibinfo {author} {\bibfnamefont {M.}~\bibnamefont {{Lassaut}}}, \bibinfo {author} {\bibfnamefont {H.}~\bibnamefont {{Flocard}}}, \bibinfo {author} {\bibfnamefont {P.}~\bibnamefont {{Bonche}}}, \bibinfo {author} {\bibfnamefont {P.~H.}\ \bibnamefont {{Heenen}}},\ and\ \bibinfo {author} {\bibfnamefont {E.}~\bibnamefont {{Suraud}}},\ }\href@noop {} {\bibfield  {journal} {\bibinfo  {journal} {Astron. Astrophys.}\ }\textbf {\bibinfo {volume} {183}},\ \bibinfo {pages} {L3} (\bibinfo {year} {1987})}\BibitemShut {NoStop}%
\bibitem [{\citenamefont {Bonche}\ \emph {et~al.}(1984)\citenamefont {Bonche}, \citenamefont {Levit},\ and\ \citenamefont {Vautherin}}]{bonche1984}%
  \BibitemOpen
  \bibfield  {author} {\bibinfo {author} {\bibfnamefont {P.}~\bibnamefont {Bonche}}, \bibinfo {author} {\bibfnamefont {S.}~\bibnamefont {Levit}},\ and\ \bibinfo {author} {\bibfnamefont {D.}~\bibnamefont {Vautherin}},\ }\href@noop {} {\bibfield  {journal} {\bibinfo  {journal} {Nuclear Physics A}\ }\textbf {\bibinfo {volume} {427}},\ \bibinfo {pages} {278} (\bibinfo {year} {1984})}\BibitemShut {NoStop}%
\bibitem [{\citenamefont {Potekhin}\ \emph {et~al.}(2015)\citenamefont {Potekhin}, \citenamefont {Pons},\ and\ \citenamefont {Page}}]{potekhin2015}%
  \BibitemOpen
  \bibfield  {author} {\bibinfo {author} {\bibfnamefont {A.~Y.}\ \bibnamefont {Potekhin}}, \bibinfo {author} {\bibfnamefont {J.~A.}\ \bibnamefont {Pons}},\ and\ \bibinfo {author} {\bibfnamefont {D.}~\bibnamefont {Page}},\ }\href {https://doi.org/10.1007/s11214-015-0180-9} {\bibfield  {journal} {\bibinfo  {journal} {Space Sci Rev}\ }\textbf {\bibinfo {volume} {191}},\ \bibinfo {pages} {239} (\bibinfo {year} {2015})}\BibitemShut {NoStop}%
\bibitem [{\citenamefont {Sandulescu}(2004)}]{sandulescu2004}%
  \BibitemOpen
  \bibfield  {author} {\bibinfo {author} {\bibfnamefont {N.}~\bibnamefont {Sandulescu}},\ }\href {https://link.aps.org/doi/10.1103/PhysRevC.70.025801} {\bibfield  {journal} {\bibinfo  {journal} {Phys. Rev. C}\ }\textbf {\bibinfo {volume} {70}},\ \bibinfo {pages} {025801} (\bibinfo {year} {2004})}\BibitemShut {NoStop}%
\bibitem [{\citenamefont {Pastore}(2012)}]{pastore2012}%
  \BibitemOpen
  \bibfield  {author} {\bibinfo {author} {\bibfnamefont {A.}~\bibnamefont {Pastore}},\ }\href {https://link.aps.org/doi/10.1103/PhysRevC.86.065802} {\bibfield  {journal} {\bibinfo  {journal} {Phys. Rev. C}\ }\textbf {\bibinfo {volume} {86}},\ \bibinfo {pages} {065802} (\bibinfo {year} {2012})}\BibitemShut {NoStop}%
\bibitem [{\citenamefont {Chamel}\ \emph {et~al.}(2013)\citenamefont {Chamel}, \citenamefont {Page},\ and\ \citenamefont {Reddy}}]{chamel2013b}%
  \BibitemOpen
  \bibfield  {author} {\bibinfo {author} {\bibfnamefont {N.}~\bibnamefont {Chamel}}, \bibinfo {author} {\bibfnamefont {D.}~\bibnamefont {Page}},\ and\ \bibinfo {author} {\bibfnamefont {S.}~\bibnamefont {Reddy}},\ }\href {https://link.aps.org/doi/10.1103/PhysRevC.87.035803} {\bibfield  {journal} {\bibinfo  {journal} {Phys. Rev. C}\ }\textbf {\bibinfo {volume} {87}},\ \bibinfo {pages} {035803} (\bibinfo {year} {2013})}\BibitemShut {NoStop}%
\bibitem [{\citenamefont {Baiko}\ \emph {et~al.}(2001)\citenamefont {Baiko}, \citenamefont {Haensel},\ and\ \citenamefont {Yakovlev}}]{baiko2001}%
  \BibitemOpen
  \bibfield  {author} {\bibinfo {author} {\bibfnamefont {D.~A.}\ \bibnamefont {Baiko}}, \bibinfo {author} {\bibfnamefont {P.}~\bibnamefont {Haensel}},\ and\ \bibinfo {author} {\bibfnamefont {D.~G.}\ \bibnamefont {Yakovlev}},\ }\href {https://www.aanda.org/articles/aa/abs/2001/28/aah2623/aah2623.html} {\bibfield  {journal} {\bibinfo  {journal} {A\&A}\ }\textbf {\bibinfo {volume} {374}},\ \bibinfo {pages} {151} (\bibinfo {year} {2001})}\BibitemShut {NoStop}%
\bibitem [{\citenamefont {Flowers}\ and\ \citenamefont {Itoh}(1976)}]{flowers1976}%
  \BibitemOpen
  \bibfield  {author} {\bibinfo {author} {\bibfnamefont {E.}~\bibnamefont {Flowers}}\ and\ \bibinfo {author} {\bibfnamefont {N.}~\bibnamefont {Itoh}},\ }\href {https://ui.adsabs.harvard.edu/abs/1976ApJ...206..218F} {\bibfield  {journal} {\bibinfo  {journal} {The Astrophysical Journal}\ }\textbf {\bibinfo {volume} {206}},\ \bibinfo {pages} {218} (\bibinfo {year} {1976})}\BibitemShut {NoStop}%
\bibitem [{\citenamefont {Flowers}\ \emph {et~al.}(1976)\citenamefont {Flowers}, \citenamefont {Ruderman},\ and\ \citenamefont {Sutherland}}]{flowers1976a}%
  \BibitemOpen
  \bibfield  {author} {\bibinfo {author} {\bibfnamefont {E.}~\bibnamefont {Flowers}}, \bibinfo {author} {\bibfnamefont {M.}~\bibnamefont {Ruderman}},\ and\ \bibinfo {author} {\bibfnamefont {P.}~\bibnamefont {Sutherland}},\ }\href {https://ui.adsabs.harvard.edu/abs/1976ApJ...205..541F} {\bibfield  {journal} {\bibinfo  {journal} {The Astrophysical Journal}\ }\textbf {\bibinfo {volume} {205}},\ \bibinfo {pages} {541} (\bibinfo {year} {1976})}\BibitemShut {NoStop}%
\bibitem [{\citenamefont {Voskresensky}\ and\ \citenamefont {Senatorov}(1987)}]{Voskresensky1987}%
  \BibitemOpen
  \bibfield  {author} {\bibinfo {author} {\bibfnamefont {D.~N.}\ \bibnamefont {Voskresensky}}\ and\ \bibinfo {author} {\bibfnamefont {A.~V.}\ \bibnamefont {Senatorov}},\ }\href@noop {} {\bibfield  {journal} {\bibinfo  {journal} {Sov. J. Nucl. Phys.}\ }\textbf {\bibinfo {volume} {45}},\ \bibinfo {pages} {411} (\bibinfo {year} {1987})}\BibitemShut {NoStop}%
\bibitem [{\citenamefont {Leinson}(2009)}]{leinson2009}%
  \BibitemOpen
  \bibfield  {author} {\bibinfo {author} {\bibfnamefont {L.~B.}\ \bibnamefont {Leinson}},\ }\href {https://link.aps.org/doi/10.1103/PhysRevC.79.045502} {\bibfield  {journal} {\bibinfo  {journal} {Phys. Rev. C}\ }\textbf {\bibinfo {volume} {79}},\ \bibinfo {pages} {045502} (\bibinfo {year} {2009})}\BibitemShut {NoStop}%
\bibitem [{\citenamefont {Leinson}(2010)}]{leinson2010}%
  \BibitemOpen
  \bibfield  {author} {\bibinfo {author} {\bibfnamefont {L.~B.}\ \bibnamefont {Leinson}},\ }\href {https://link.aps.org/doi/10.1103/PhysRevC.81.025501} {\bibfield  {journal} {\bibinfo  {journal} {Phys. Rev. C}\ }\textbf {\bibinfo {volume} {81}},\ \bibinfo {pages} {025501} (\bibinfo {year} {2010})}\BibitemShut {NoStop}%
\bibitem [{\citenamefont {Makishima}\ \emph {et~al.}(2014)\citenamefont {Makishima}, \citenamefont {Enoto}, \citenamefont {Hiraga}, \citenamefont {Nakano}, \citenamefont {Nakazawa}, \citenamefont {Sakurai}, \citenamefont {Sasano},\ and\ \citenamefont {Murakami}}]{makishima2014}%
  \BibitemOpen
  \bibfield  {author} {\bibinfo {author} {\bibfnamefont {K.}~\bibnamefont {Makishima}}, \bibinfo {author} {\bibfnamefont {T.}~\bibnamefont {Enoto}}, \bibinfo {author} {\bibfnamefont {J.~S.}\ \bibnamefont {Hiraga}}, \bibinfo {author} {\bibfnamefont {T.}~\bibnamefont {Nakano}}, \bibinfo {author} {\bibfnamefont {K.}~\bibnamefont {Nakazawa}}, \bibinfo {author} {\bibfnamefont {S.}~\bibnamefont {Sakurai}}, \bibinfo {author} {\bibfnamefont {M.}~\bibnamefont {Sasano}},\ and\ \bibinfo {author} {\bibfnamefont {H.}~\bibnamefont {Murakami}},\ }\href {https://link.aps.org/doi/10.1103/PhysRevLett.112.171102} {\bibfield  {journal} {\bibinfo  {journal} {Phys. Rev. Lett.}\ }\textbf {\bibinfo {volume} {112}},\ \bibinfo {pages} {171102} (\bibinfo {year} {2014})}\BibitemShut {NoStop}%
\bibitem [{\citenamefont {Turolla}\ \emph {et~al.}(2015)\citenamefont {Turolla}, \citenamefont {Zane},\ and\ \citenamefont {Watts}}]{turolla2015}%
  \BibitemOpen
  \bibfield  {author} {\bibinfo {author} {\bibfnamefont {R.}~\bibnamefont {Turolla}}, \bibinfo {author} {\bibfnamefont {S.}~\bibnamefont {Zane}},\ and\ \bibinfo {author} {\bibfnamefont {A.~L.}\ \bibnamefont {Watts}},\ }\href {https://ui.adsabs.harvard.edu/abs/2015RPPh...78k6901T} {\bibfield  {journal} {\bibinfo  {journal} {Rep. Prog. Phys.}\ }\textbf {\bibinfo {volume} {78}},\ \bibinfo {pages} {116901} (\bibinfo {year} {2015})}\BibitemShut {NoStop}%
\bibitem [{\citenamefont {Kaspi}\ and\ \citenamefont {Beloborodov}(2017)}]{kaspi2017}%
  \BibitemOpen
  \bibfield  {author} {\bibinfo {author} {\bibfnamefont {V.~M.}\ \bibnamefont {Kaspi}}\ and\ \bibinfo {author} {\bibfnamefont {A.~M.}\ \bibnamefont {Beloborodov}},\ }\href {https://www.annualreviews.org/content/journals/10.1146/annurev-astro-081915-023329} {\bibfield  {journal} {\bibinfo  {journal} {Annu. Rev. Astron. Astrophys.}\ }\textbf {\bibinfo {volume} {55}},\ \bibinfo {pages} {261} (\bibinfo {year} {2017})}\BibitemShut {NoStop}%
\bibitem [{\citenamefont {Esposito}\ \emph {et~al.}(2021)\citenamefont {Esposito}, \citenamefont {Rea},\ and\ \citenamefont {Israel}}]{esposito2021}%
  \BibitemOpen
  \bibfield  {author} {\bibinfo {author} {\bibfnamefont {P.}~\bibnamefont {Esposito}}, \bibinfo {author} {\bibfnamefont {N.}~\bibnamefont {Rea}},\ and\ \bibinfo {author} {\bibfnamefont {G.~L.}\ \bibnamefont {Israel}},\ }in\ \href {https://doi.org/10.1007/978-3-662-62110-3_3} {\emph {\bibinfo {booktitle} {Timing {{Neutron Stars}}: {{Pulsations}}, {{Oscillations}} and {{Explosions}}}}},\ \bibinfo {editor} {edited by\ \bibinfo {editor} {\bibfnamefont {T.~M.}\ \bibnamefont {Belloni}}, \bibinfo {editor} {\bibfnamefont {M.}~\bibnamefont {M{\'e}ndez}},\ and\ \bibinfo {editor} {\bibfnamefont {C.}~\bibnamefont {Zhang}}}\ (\bibinfo  {publisher} {Springer},\ \bibinfo {address} {Berlin, Heidelberg},\ \bibinfo {year} {2021})\ pp.\ \bibinfo {pages} {97--142}\BibitemShut {NoStop}%
\bibitem [{\citenamefont {Pe\~na Arteaga}\ \emph {et~al.}(2011)\citenamefont {Pe\~na Arteaga}, \citenamefont {Grasso}, \citenamefont {Khan},\ and\ \citenamefont {Ring}}]{arteaga2011}%
  \BibitemOpen
  \bibfield  {author} {\bibinfo {author} {\bibfnamefont {D.}~\bibnamefont {Pe\~na Arteaga}}, \bibinfo {author} {\bibfnamefont {M.}~\bibnamefont {Grasso}}, \bibinfo {author} {\bibfnamefont {E.}~\bibnamefont {Khan}},\ and\ \bibinfo {author} {\bibfnamefont {P.}~\bibnamefont {Ring}},\ }\href {https://doi.org/10.1103/PhysRevC.84.045806} {\bibfield  {journal} {\bibinfo  {journal} {Phys. Rev. C}\ }\textbf {\bibinfo {volume} {84}},\ \bibinfo {pages} {045806} (\bibinfo {year} {2011})}\BibitemShut {NoStop}%
\bibitem [{\citenamefont {Chamel}\ \emph {et~al.}(2012)\citenamefont {Chamel}, \citenamefont {Pavlov}, \citenamefont {Mihailov}, \citenamefont {Velchev}, \citenamefont {Stoyanov}, \citenamefont {Mutafchieva}, \citenamefont {Ivanovich}, \citenamefont {Pearson},\ and\ \citenamefont {Goriely}}]{chamel2012a}%
  \BibitemOpen
  \bibfield  {author} {\bibinfo {author} {\bibfnamefont {N.}~\bibnamefont {Chamel}}, \bibinfo {author} {\bibfnamefont {R.~L.}\ \bibnamefont {Pavlov}}, \bibinfo {author} {\bibfnamefont {L.~M.}\ \bibnamefont {Mihailov}}, \bibinfo {author} {\bibfnamefont {{\relax Ch}.~J.}\ \bibnamefont {Velchev}}, \bibinfo {author} {\bibfnamefont {{\relax Zh}.~K.}\ \bibnamefont {Stoyanov}}, \bibinfo {author} {\bibfnamefont {Y.~D.}\ \bibnamefont {Mutafchieva}}, \bibinfo {author} {\bibfnamefont {M.~D.}\ \bibnamefont {Ivanovich}}, \bibinfo {author} {\bibfnamefont {J.~M.}\ \bibnamefont {Pearson}},\ and\ \bibinfo {author} {\bibfnamefont {S.}~\bibnamefont {Goriely}},\ }\href {https://link.aps.org/doi/10.1103/PhysRevC.86.055804} {\bibfield  {journal} {\bibinfo  {journal} {Phys. Rev. C}\ }\textbf {\bibinfo {volume} {86}},\ \bibinfo {pages} {055804} (\bibinfo {year} {2012})}\BibitemShut {NoStop}%
\bibitem [{\citenamefont {Basilico}\ \emph {et~al.}(2015)\citenamefont {Basilico}, \citenamefont {Arteaga}, \citenamefont {{Roca-Maza}},\ and\ \citenamefont {Col{\`o}}}]{basilico2015}%
  \BibitemOpen
  \bibfield  {author} {\bibinfo {author} {\bibfnamefont {D.}~\bibnamefont {Basilico}}, \bibinfo {author} {\bibfnamefont {D.~P.}\ \bibnamefont {Arteaga}}, \bibinfo {author} {\bibfnamefont {X.}~\bibnamefont {{Roca-Maza}}},\ and\ \bibinfo {author} {\bibfnamefont {G.}~\bibnamefont {Col{\`o}}},\ }\href {https://link.aps.org/doi/10.1103/PhysRevC.92.035802} {\bibfield  {journal} {\bibinfo  {journal} {Phys. Rev. C}\ }\textbf {\bibinfo {volume} {92}},\ \bibinfo {pages} {035802} (\bibinfo {year} {2015})}\BibitemShut {NoStop}%
\bibitem [{\citenamefont {Parmar}\ \emph {et~al.}(2023)\citenamefont {Parmar}, \citenamefont {Das}, \citenamefont {Sharma},\ and\ \citenamefont {Patra}}]{parmar2023a}%
  \BibitemOpen
  \bibfield  {author} {\bibinfo {author} {\bibfnamefont {V.}~\bibnamefont {Parmar}}, \bibinfo {author} {\bibfnamefont {H.~C.}\ \bibnamefont {Das}}, \bibinfo {author} {\bibfnamefont {M.~K.}\ \bibnamefont {Sharma}},\ and\ \bibinfo {author} {\bibfnamefont {S.~K.}\ \bibnamefont {Patra}},\ }\href {https://link.aps.org/doi/10.1103/PhysRevD.107.043022} {\bibfield  {journal} {\bibinfo  {journal} {Phys. Rev. D}\ }\textbf {\bibinfo {volume} {107}},\ \bibinfo {pages} {043022} (\bibinfo {year} {2023})}\BibitemShut {NoStop}%
\bibitem [{\citenamefont {Sekizawa}\ and\ \citenamefont {Kaba}(2023)}]{sekizawa2023a}%
  \BibitemOpen
  \bibfield  {author} {\bibinfo {author} {\bibfnamefont {K.}~\bibnamefont {Sekizawa}}\ and\ \bibinfo {author} {\bibfnamefont {K.}~\bibnamefont {Kaba}},\ }\href {http://arxiv.org/abs/2302.07923} {\bibinfo {title} {Possibleexistence of extremely neutron-rich superheavy nuclei in neutron star crusts under a superstrong magnetic field}} (\bibinfo {year} {2023}),\ \Eprint {https://arxiv.org/abs/2302.07923} {arXiv:2302.07923 [nucl-th]} \BibitemShut {NoStop}%
\bibitem [{\citenamefont {Bonanno}\ \emph {et~al.}(2003)\citenamefont {Bonanno}, \citenamefont {Rezzolla},\ and\ \citenamefont {Urpin}}]{bonanno2003}%
  \BibitemOpen
  \bibfield  {author} {\bibinfo {author} {\bibfnamefont {A.}~\bibnamefont {Bonanno}}, \bibinfo {author} {\bibfnamefont {L.}~\bibnamefont {Rezzolla}},\ and\ \bibinfo {author} {\bibfnamefont {V.}~\bibnamefont {Urpin}},\ }\href {https://www.aanda.org/articles/aa/abs/2003/42/aafg163/aafg163.html} {\bibfield  {journal} {\bibinfo  {journal} {A\&A}\ }\textbf {\bibinfo {volume} {410}},\ \bibinfo {pages} {L33} (\bibinfo {year} {2003})}\BibitemShut {NoStop}%
\bibitem [{\citenamefont {Naso}\ \emph {et~al.}(2008)\citenamefont {Naso}, \citenamefont {Rezzolla}, \citenamefont {Bonanno},\ and\ \citenamefont {Patern{\`o}}}]{naso2008}%
  \BibitemOpen
  \bibfield  {author} {\bibinfo {author} {\bibfnamefont {L.}~\bibnamefont {Naso}}, \bibinfo {author} {\bibfnamefont {L.}~\bibnamefont {Rezzolla}}, \bibinfo {author} {\bibfnamefont {A.}~\bibnamefont {Bonanno}},\ and\ \bibinfo {author} {\bibfnamefont {L.}~\bibnamefont {Patern{\`o}}},\ }\href {http://www.aanda.org/10.1051/0004-6361:20078360} {\bibfield  {journal} {\bibinfo  {journal} {A\&A}\ }\textbf {\bibinfo {volume} {479}},\ \bibinfo {pages} {167} (\bibinfo {year} {2008})}\BibitemShut {NoStop}%
\bibitem [{\citenamefont {Frieben}\ and\ \citenamefont {Rezzolla}(2012)}]{frieben2012}%
  \BibitemOpen
  \bibfield  {author} {\bibinfo {author} {\bibfnamefont {J.}~\bibnamefont {Frieben}}\ and\ \bibinfo {author} {\bibfnamefont {L.}~\bibnamefont {Rezzolla}},\ }\href {https://doi.org/10.1111/j.1365-2966.2012.22027.x} {\bibfield  {journal} {\bibinfo  {journal} {Monthly Notices of the Royal Astronomical Society}\ }\textbf {\bibinfo {volume} {427}},\ \bibinfo {pages} {3406} (\bibinfo {year} {2012})}\BibitemShut {NoStop}%
\bibitem [{\citenamefont {Potekhin}\ and\ \citenamefont {Yakovlev}(1996)}]{potekhin1996}%
  \BibitemOpen
  \bibfield  {author} {\bibinfo {author} {\bibfnamefont {A.~Y.}\ \bibnamefont {Potekhin}}\ and\ \bibinfo {author} {\bibfnamefont {D.~G.}\ \bibnamefont {Yakovlev}},\ }\href {https://ui.adsabs.harvard.edu/abs/1996A&A...314..341P} {\bibinfo {title} {Electron conduction along quantizing magnetic fields in neutron star crusts. {{II}}. {{Practical}} formulae.}} (\bibinfo {year} {1996})\BibitemShut {NoStop}%
\bibitem [{\citenamefont {Potekhin}(1999)}]{potekhin1999}%
  \BibitemOpen
  \bibfield  {author} {\bibinfo {author} {\bibfnamefont {A.~Y.}\ \bibnamefont {Potekhin}},\ }\href {https://ui.adsabs.harvard.edu/abs/1999A&A...351..787P} {\bibinfo {title} {Electron conduction in magnetized neutron star envelopes}} (\bibinfo {year} {1999})\BibitemShut {NoStop}%
\bibitem [{\citenamefont {Broderick}\ \emph {et~al.}(2000)\citenamefont {Broderick}, \citenamefont {Prakash},\ and\ \citenamefont {Lattimer}}]{broderick2000}%
  \BibitemOpen
  \bibfield  {author} {\bibinfo {author} {\bibfnamefont {A.}~\bibnamefont {Broderick}}, \bibinfo {author} {\bibfnamefont {M.}~\bibnamefont {Prakash}},\ and\ \bibinfo {author} {\bibfnamefont {J.~M.}\ \bibnamefont {Lattimer}},\ }\href {https://doi.org/10.1086/309010} {\bibfield  {journal} {\bibinfo  {journal} {The Astrophysical Journal}\ }\textbf {\bibinfo {volume} {537}},\ \bibinfo {pages} {351} (\bibinfo {year} {2000})}\BibitemShut {NoStop}%
\bibitem [{\citenamefont {Ventura}\ and\ \citenamefont {Potekhin}(2001)}]{Ventura:2001br}%
  \BibitemOpen
  \bibfield  {author} {\bibinfo {author} {\bibfnamefont {J.}~\bibnamefont {Ventura}}\ and\ \bibinfo {author} {\bibfnamefont {A.~Y.}\ \bibnamefont {Potekhin}},\ }\href {https://arxiv.org/abs/astro-ph/0104003} {\bibinfo {title} {Neutron star envelopes and thermal radiation from the magnetic surface}} (\bibinfo {year} {2001}),\ \Eprint {https://arxiv.org/abs/astro-ph/0104003} {arXiv:astro-ph/0104003 [astro-ph]} \BibitemShut {NoStop}%
\bibitem [{\citenamefont {Stein}\ \emph {et~al.}(2016)\citenamefont {Stein}, \citenamefont {Maruhn}, \citenamefont {Sedrakian},\ and\ \citenamefont {Reinhard}}]{stein2016}%
  \BibitemOpen
  \bibfield  {author} {\bibinfo {author} {\bibfnamefont {M.}~\bibnamefont {Stein}}, \bibinfo {author} {\bibfnamefont {J.}~\bibnamefont {Maruhn}}, \bibinfo {author} {\bibfnamefont {A.}~\bibnamefont {Sedrakian}},\ and\ \bibinfo {author} {\bibfnamefont {P.-G.}\ \bibnamefont {Reinhard}},\ }\href {https://link.aps.org/doi/10.1103/PhysRevC.94.035802} {\bibfield  {journal} {\bibinfo  {journal} {Phys. Rev. C}\ }\textbf {\bibinfo {volume} {94}},\ \bibinfo {pages} {035802} (\bibinfo {year} {2016})}\BibitemShut {NoStop}%
\bibitem [{\citenamefont {Jiang}\ and\ \citenamefont {Chen}(2024)}]{jiang2024}%
  \BibitemOpen
  \bibfield  {author} {\bibinfo {author} {\bibfnamefont {W.}~\bibnamefont {Jiang}}\ and\ \bibinfo {author} {\bibfnamefont {Y.-j.}\ \bibnamefont {Chen}},\ }\href {https://dx.doi.org/10.1088/1674-1137/ad39cc} {\bibfield  {journal} {\bibinfo  {journal} {Chinese Phys. C}\ }\textbf {\bibinfo {volume} {48}},\ \bibinfo {pages} {074103} (\bibinfo {year} {2024})}\BibitemShut {NoStop}%
\bibitem [{\citenamefont {Basilico}\ \emph {et~al.}(2024)\citenamefont {Basilico}, \citenamefont {{Roca-Maza}},\ and\ \citenamefont {Col{\`o}}}]{basilico2024}%
  \BibitemOpen
  \bibfield  {author} {\bibinfo {author} {\bibfnamefont {D.}~\bibnamefont {Basilico}}, \bibinfo {author} {\bibfnamefont {X.}~\bibnamefont {{Roca-Maza}}},\ and\ \bibinfo {author} {\bibfnamefont {G.}~\bibnamefont {Col{\`o}}},\ }\href {https://doi.org/10.48550/arXiv.2403.17773} {\bibfield  {journal} {\bibinfo  {journal} {arXiv:2403.17773 [nucl-th]}\ } (\bibinfo {year} {2024})}\BibitemShut {NoStop}%
\bibitem [{\citenamefont {Tajima}\ \emph {et~al.}(2023)\citenamefont {Tajima}, \citenamefont {Funaki}, \citenamefont {Sekino}, \citenamefont {Yasutake},\ and\ \citenamefont {Matsuo}}]{tajima2023}%
  \BibitemOpen
  \bibfield  {author} {\bibinfo {author} {\bibfnamefont {H.}~\bibnamefont {Tajima}}, \bibinfo {author} {\bibfnamefont {H.}~\bibnamefont {Funaki}}, \bibinfo {author} {\bibfnamefont {Y.}~\bibnamefont {Sekino}}, \bibinfo {author} {\bibfnamefont {N.}~\bibnamefont {Yasutake}},\ and\ \bibinfo {author} {\bibfnamefont {M.}~\bibnamefont {Matsuo}},\ }\href {https://doi.org/10.1103/PhysRevC.108.L052802} {\bibfield  {journal} {\bibinfo  {journal} {Phys. Rev. C}\ }\textbf {\bibinfo {volume} {108}},\ \bibinfo {pages} {L052802} (\bibinfo {year} {2023})}\BibitemShut {NoStop}%
\bibitem [{\citenamefont {Fulde}\ and\ \citenamefont {Ferrell}(1964)}]{fulde1964}%
  \BibitemOpen
  \bibfield  {author} {\bibinfo {author} {\bibfnamefont {P.}~\bibnamefont {Fulde}}\ and\ \bibinfo {author} {\bibfnamefont {R.~A.}\ \bibnamefont {Ferrell}},\ }\href {https://link.aps.org/doi/10.1103/PhysRev.135.A550} {\bibfield  {journal} {\bibinfo  {journal} {Phys. Rev.}\ }\textbf {\bibinfo {volume} {135}},\ \bibinfo {pages} {A550} (\bibinfo {year} {1964})}\BibitemShut {NoStop}%
\bibitem [{\citenamefont {Larkin}\ and\ \citenamefont {Ovchinnikov}(1964)}]{larkin1964}%
  \BibitemOpen
  \bibfield  {author} {\bibinfo {author} {\bibfnamefont {A.~I.}\ \bibnamefont {Larkin}}\ and\ \bibinfo {author} {\bibfnamefont {Y.~N.}\ \bibnamefont {Ovchinnikov}},\ }\href@noop {} {\bibfield  {journal} {\bibinfo  {journal} {Zh. Eksp. Teor. Fiz.}\ }\textbf {\bibinfo {volume} {47}},\ \bibinfo {pages} {1136} (\bibinfo {year} {1964})}\BibitemShut {NoStop}%
\bibitem [{\citenamefont {Casalbuoni}\ and\ \citenamefont {Nardulli}(2004)}]{casalbuoni2004}%
  \BibitemOpen
  \bibfield  {author} {\bibinfo {author} {\bibfnamefont {R.}~\bibnamefont {Casalbuoni}}\ and\ \bibinfo {author} {\bibfnamefont {G.}~\bibnamefont {Nardulli}},\ }\href {https://link.aps.org/doi/10.1103/RevModPhys.76.263} {\bibfield  {journal} {\bibinfo  {journal} {Rev. Mod. Phys.}\ }\textbf {\bibinfo {volume} {76}},\ \bibinfo {pages} {263} (\bibinfo {year} {2004})}\BibitemShut {NoStop}%
\bibitem [{\citenamefont {Magierski}\ \emph {et~al.}(2019)\citenamefont {Magierski}, \citenamefont {T{\"u}zemen},\ and\ \citenamefont {Wlaz{\l}owski}}]{magierski2019}%
  \BibitemOpen
  \bibfield  {author} {\bibinfo {author} {\bibfnamefont {P.}~\bibnamefont {Magierski}}, \bibinfo {author} {\bibfnamefont {B.}~\bibnamefont {T{\"u}zemen}},\ and\ \bibinfo {author} {\bibfnamefont {G.}~\bibnamefont {Wlaz{\l}owski}},\ }\href {https://link.aps.org/doi/10.1103/PhysRevA.100.033613} {\bibfield  {journal} {\bibinfo  {journal} {Phys. Rev. A}\ }\textbf {\bibinfo {volume} {100}},\ \bibinfo {pages} {033613} (\bibinfo {year} {2019})}\BibitemShut {NoStop}%
\bibitem [{\citenamefont {Magierski}\ \emph {et~al.}(2021)\citenamefont {Magierski}, \citenamefont {T{\"u}zemen},\ and\ \citenamefont {Wlaz{\l}owski}}]{magierski2021}%
  \BibitemOpen
  \bibfield  {author} {\bibinfo {author} {\bibfnamefont {P.}~\bibnamefont {Magierski}}, \bibinfo {author} {\bibfnamefont {B.}~\bibnamefont {T{\"u}zemen}},\ and\ \bibinfo {author} {\bibfnamefont {G.}~\bibnamefont {Wlaz{\l}owski}},\ }\href {https://link.aps.org/doi/10.1103/PhysRevA.104.033304} {\bibfield  {journal} {\bibinfo  {journal} {Phys. Rev. A}\ }\textbf {\bibinfo {volume} {104}},\ \bibinfo {pages} {033304} (\bibinfo {year} {2021})}\BibitemShut {NoStop}%
\bibitem [{\citenamefont {T{\"u}zemen}\ \emph {et~al.}(2023)\citenamefont {T{\"u}zemen}, \citenamefont {Zawi{\'s}lak}, \citenamefont {Wlaz{\l}owski},\ and\ \citenamefont {Magierski}}]{tuzemen2023}%
  \BibitemOpen
  \bibfield  {author} {\bibinfo {author} {\bibfnamefont {B.}~\bibnamefont {T{\"u}zemen}}, \bibinfo {author} {\bibfnamefont {T.}~\bibnamefont {Zawi{\'s}lak}}, \bibinfo {author} {\bibfnamefont {G.}~\bibnamefont {Wlaz{\l}owski}},\ and\ \bibinfo {author} {\bibfnamefont {P.}~\bibnamefont {Magierski}},\ }\href {https://dx.doi.org/10.1088/1367-2630/acc26b} {\bibfield  {journal} {\bibinfo  {journal} {New J. Phys.}\ }\textbf {\bibinfo {volume} {25}},\ \bibinfo {pages} {033013} (\bibinfo {year} {2023})}\BibitemShut {NoStop}%
\bibitem [{\citenamefont {Bogdanov}\ and\ \citenamefont {R{\"o}{\ss}ler}(2001)}]{bogdanov2001}%
  \BibitemOpen
  \bibfield  {author} {\bibinfo {author} {\bibfnamefont {A.~N.}\ \bibnamefont {Bogdanov}}\ and\ \bibinfo {author} {\bibfnamefont {U.~K.}\ \bibnamefont {R{\"o}{\ss}ler}},\ }\href {https://link.aps.org/doi/10.1103/PhysRevLett.87.037203} {\bibfield  {journal} {\bibinfo  {journal} {Phys. Rev. Lett.}\ }\textbf {\bibinfo {volume} {87}},\ \bibinfo {pages} {037203} (\bibinfo {year} {2001})}\BibitemShut {NoStop}%
\bibitem [{\citenamefont {Zhou}\ \emph {et~al.}(2015)\citenamefont {Zhou}, \citenamefont {Iacocca}, \citenamefont {Awad}, \citenamefont {Dumas}, \citenamefont {Zhang}, \citenamefont {Braun},\ and\ \citenamefont {{\AA}kerman}}]{zhou2015}%
  \BibitemOpen
  \bibfield  {author} {\bibinfo {author} {\bibfnamefont {Y.}~\bibnamefont {Zhou}}, \bibinfo {author} {\bibfnamefont {E.}~\bibnamefont {Iacocca}}, \bibinfo {author} {\bibfnamefont {A.~A.}\ \bibnamefont {Awad}}, \bibinfo {author} {\bibfnamefont {R.~K.}\ \bibnamefont {Dumas}}, \bibinfo {author} {\bibfnamefont {F.~C.}\ \bibnamefont {Zhang}}, \bibinfo {author} {\bibfnamefont {H.~B.}\ \bibnamefont {Braun}},\ and\ \bibinfo {author} {\bibfnamefont {J.}~\bibnamefont {{\AA}kerman}},\ }\href {https://www.nature.com/articles/ncomms9193} {\bibfield  {journal} {\bibinfo  {journal} {Nat Commun}\ }\textbf {\bibinfo {volume} {6}},\ \bibinfo {pages} {8193} (\bibinfo {year} {2015})}\BibitemShut {NoStop}%
\bibitem [{\citenamefont {G^^c3^^b6bel}\ \emph {et~al.}(2021)\citenamefont {G^^c3^^b6bel}, \citenamefont {Mertig},\ and\ \citenamefont {Tretiakov}}]{gobel2021}%
  \BibitemOpen
  \bibfield  {author} {\bibinfo {author} {\bibfnamefont {B.}~\bibnamefont {G^^c3^^b6bel}}, \bibinfo {author} {\bibfnamefont {I.}~\bibnamefont {Mertig}},\ and\ \bibinfo {author} {\bibfnamefont {O.~A.}\ \bibnamefont {Tretiakov}},\ }\href {https://doi.org/https://doi.org/10.1016/j.physrep.2020.10.001} {\bibfield  {journal} {\bibinfo  {journal} {Physics Reports}\ }\textbf {\bibinfo {volume} {895}},\ \bibinfo {pages} {1} (\bibinfo {year} {2021})},\ \bibinfo {note} {beyond skyrmions: Review and perspectives of alternative magnetic quasiparticles}\BibitemShut {NoStop}%
\bibitem [{\citenamefont {Bulgac}\ and\ \citenamefont {Yu}(2002)}]{bulgac2002}%
  \BibitemOpen
  \bibfield  {author} {\bibinfo {author} {\bibfnamefont {A.}~\bibnamefont {Bulgac}}\ and\ \bibinfo {author} {\bibfnamefont {Y.}~\bibnamefont {Yu}},\ }\href {https://link.aps.org/doi/10.1103/PhysRevLett.88.042504} {\bibfield  {journal} {\bibinfo  {journal} {Phys. Rev. Lett.}\ }\textbf {\bibinfo {volume} {88}},\ \bibinfo {pages} {042504} (\bibinfo {year} {2002})}\BibitemShut {NoStop}%
\bibitem [{\citenamefont {Bulgac}(2002)}]{bulgac2002a}%
  \BibitemOpen
  \bibfield  {author} {\bibinfo {author} {\bibfnamefont {A.}~\bibnamefont {Bulgac}},\ }\href {https://link.aps.org/doi/10.1103/PhysRevC.65.051305} {\bibfield  {journal} {\bibinfo  {journal} {Phys. Rev. C}\ }\textbf {\bibinfo {volume} {65}},\ \bibinfo {pages} {051305} (\bibinfo {year} {2002})}\BibitemShut {NoStop}%
\bibitem [{\citenamefont {Jin}\ \emph {et~al.}(2021)\citenamefont {Jin}, \citenamefont {Roche}, \citenamefont {Stetcu}, \citenamefont {Abdurrahman},\ and\ \citenamefont {Bulgac}}]{jin2021}%
  \BibitemOpen
  \bibfield  {author} {\bibinfo {author} {\bibfnamefont {S.}~\bibnamefont {Jin}}, \bibinfo {author} {\bibfnamefont {K.~J.}\ \bibnamefont {Roche}}, \bibinfo {author} {\bibfnamefont {I.}~\bibnamefont {Stetcu}}, \bibinfo {author} {\bibfnamefont {I.}~\bibnamefont {Abdurrahman}},\ and\ \bibinfo {author} {\bibfnamefont {A.}~\bibnamefont {Bulgac}},\ }\href {https://www.sciencedirect.com/science/article/pii/S0010465521002423} {\bibfield  {journal} {\bibinfo  {journal} {Computer Physics Communications}\ }\textbf {\bibinfo {volume} {269}},\ \bibinfo {pages} {108130} (\bibinfo {year} {2021})}\BibitemShut {NoStop}%
\bibitem [{\citenamefont {Goodman}(1981)}]{goodman1981}%
  \BibitemOpen
  \bibfield  {author} {\bibinfo {author} {\bibfnamefont {A.~L.}\ \bibnamefont {Goodman}},\ }\href {https://www.sciencedirect.com/science/article/pii/0375947481905571} {\bibfield  {journal} {\bibinfo  {journal} {Nuclear Physics A}\ }\textbf {\bibinfo {volume} {352}},\ \bibinfo {pages} {30} (\bibinfo {year} {1981})}\BibitemShut {NoStop}%
\bibitem [{\citenamefont {Duguet}\ and\ \citenamefont {Ryssens}(2020)}]{duguet2020}%
  \BibitemOpen
  \bibfield  {author} {\bibinfo {author} {\bibfnamefont {T.}~\bibnamefont {Duguet}}\ and\ \bibinfo {author} {\bibfnamefont {W.}~\bibnamefont {Ryssens}},\ }\href {https://link.aps.org/doi/10.1103/PhysRevC.102.044328} {\bibfield  {journal} {\bibinfo  {journal} {Phys. Rev. C}\ }\textbf {\bibinfo {volume} {102}},\ \bibinfo {pages} {044328} (\bibinfo {year} {2020})}\BibitemShut {NoStop}%
\bibitem [{\citenamefont {Hellemans}\ \emph {et~al.}(2012)\citenamefont {Hellemans}, \citenamefont {Heenen},\ and\ \citenamefont {Bender}}]{Hellemans2012}%
  \BibitemOpen
  \bibfield  {author} {\bibinfo {author} {\bibfnamefont {V.}~\bibnamefont {Hellemans}}, \bibinfo {author} {\bibfnamefont {P.-H.}\ \bibnamefont {Heenen}},\ and\ \bibinfo {author} {\bibfnamefont {M.}~\bibnamefont {Bender}},\ }\href {https://doi.org/10.1103/PhysRevC.85.014326} {\bibfield  {journal} {\bibinfo  {journal} {Phys. Rev. C}\ }\textbf {\bibinfo {volume} {85}},\ \bibinfo {pages} {014326} (\bibinfo {year} {2012})}\BibitemShut {NoStop}%
\bibitem [{\citenamefont {Sekizawa}\ and\ \citenamefont {Yabana}(2013)}]{Sekizawa2013}%
  \BibitemOpen
  \bibfield  {author} {\bibinfo {author} {\bibfnamefont {K.}~\bibnamefont {Sekizawa}}\ and\ \bibinfo {author} {\bibfnamefont {K.}~\bibnamefont {Yabana}},\ }\href {https://doi.org/10.1103/PhysRevC.88.014614} {\bibfield  {journal} {\bibinfo  {journal} {Phys. Rev. C}\ }\textbf {\bibinfo {volume} {88}},\ \bibinfo {pages} {014614} (\bibinfo {year} {2013})}\BibitemShut {NoStop}%
\bibitem [{\citenamefont {Magierski}\ and\ \citenamefont {Bulgac}(2004)}]{magierski2004}%
  \BibitemOpen
  \bibfield  {author} {\bibinfo {author} {\bibfnamefont {P.}~\bibnamefont {Magierski}}\ and\ \bibinfo {author} {\bibfnamefont {A.}~\bibnamefont {Bulgac}},\ }\href {https://www.sciencedirect.com/science/article/pii/S0375947404005603} {\bibfield  {journal} {\bibinfo  {journal} {Nuclear Physics A}\ }\bibinfo {series} {Proceedings of the 8th {{International Conference}} on {{Clustering Aspects}} of {{Nuclear Structure}} and {{Dynamics}}},\ \textbf {\bibinfo {volume} {738}},\ \bibinfo {pages} {143} (\bibinfo {year} {2004})}\BibitemShut {NoStop}%
\bibitem [{\citenamefont {Martin}\ and\ \citenamefont {Urban}(2016)}]{martin2016}%
  \BibitemOpen
  \bibfield  {author} {\bibinfo {author} {\bibfnamefont {N.}~\bibnamefont {Martin}}\ and\ \bibinfo {author} {\bibfnamefont {M.}~\bibnamefont {Urban}},\ }\href {https://link.aps.org/doi/10.1103/PhysRevC.94.065801} {\bibfield  {journal} {\bibinfo  {journal} {Phys. Rev. C}\ }\textbf {\bibinfo {volume} {94}},\ \bibinfo {pages} {065801} (\bibinfo {year} {2016})}\BibitemShut {NoStop}%
\bibitem [{\citenamefont {Thi}\ \emph {et~al.}(2023)\citenamefont {Thi}, \citenamefont {Fantina},\ and\ \citenamefont {Gulminelli}}]{thi2023}%
  \BibitemOpen
  \bibfield  {author} {\bibinfo {author} {\bibfnamefont {H.~D.}\ \bibnamefont {Thi}}, \bibinfo {author} {\bibfnamefont {A.~F.}\ \bibnamefont {Fantina}},\ and\ \bibinfo {author} {\bibfnamefont {F.}~\bibnamefont {Gulminelli}},\ }\href {https://www.aanda.org/articles/aa/abs/2023/04/aa45061-22/aa45061-22.html} {\bibfield  {journal} {\bibinfo  {journal} {A\&A}\ }\textbf {\bibinfo {volume} {672}},\ \bibinfo {pages} {A160} (\bibinfo {year} {2023})}\BibitemShut {NoStop}%
\bibitem [{\citenamefont {Drissi}\ and\ \citenamefont {Rios}(2022)}]{drissi2022}%
  \BibitemOpen
  \bibfield  {author} {\bibinfo {author} {\bibfnamefont {M.}~\bibnamefont {Drissi}}\ and\ \bibinfo {author} {\bibfnamefont {A.}~\bibnamefont {Rios}},\ }\href {https://doi.org/10.1140/epja/s10050-022-00738-2} {\bibfield  {journal} {\bibinfo  {journal} {Eur. Phys. J. A}\ }\textbf {\bibinfo {volume} {58}},\ \bibinfo {pages} {90} (\bibinfo {year} {2022})}\BibitemShut {NoStop}%
\bibitem [{\citenamefont {Yu}\ and\ \citenamefont {Bulgac}(2003)}]{yu2003}%
  \BibitemOpen
  \bibfield  {author} {\bibinfo {author} {\bibfnamefont {Y.}~\bibnamefont {Yu}}\ and\ \bibinfo {author} {\bibfnamefont {A.}~\bibnamefont {Bulgac}},\ }\href {https://link.aps.org/doi/10.1103/PhysRevLett.90.222501} {\bibfield  {journal} {\bibinfo  {journal} {Phys. Rev. Lett.}\ }\textbf {\bibinfo {volume} {90}},\ \bibinfo {pages} {222501} (\bibinfo {year} {2003})}\BibitemShut {NoStop}%
\bibitem [{\citenamefont {Bulgac}\ \emph {et~al.}(2018)\citenamefont {Bulgac}, \citenamefont {Forbes}, \citenamefont {Jin}, \citenamefont {Perez},\ and\ \citenamefont {Schunck}}]{bulgac2018}%
  \BibitemOpen
  \bibfield  {author} {\bibinfo {author} {\bibfnamefont {A.}~\bibnamefont {Bulgac}}, \bibinfo {author} {\bibfnamefont {M.~M.}\ \bibnamefont {Forbes}}, \bibinfo {author} {\bibfnamefont {S.}~\bibnamefont {Jin}}, \bibinfo {author} {\bibfnamefont {R.~N.}\ \bibnamefont {Perez}},\ and\ \bibinfo {author} {\bibfnamefont {N.}~\bibnamefont {Schunck}},\ }\href {https://link.aps.org/doi/10.1103/PhysRevC.97.044313} {\bibfield  {journal} {\bibinfo  {journal} {Phys. Rev. C}\ }\textbf {\bibinfo {volume} {97}},\ \bibinfo {pages} {044313} (\bibinfo {year} {2018})}\BibitemShut {NoStop}%
\bibitem [{\citenamefont {Wlaz{\l}owski}\ \emph {et~al.}(2016)\citenamefont {Wlaz{\l}owski}, \citenamefont {Sekizawa}, \citenamefont {Magierski}, \citenamefont {Bulgac},\ and\ \citenamefont {Forbes}}]{wlazlowski2016a}%
  \BibitemOpen
  \bibfield  {author} {\bibinfo {author} {\bibfnamefont {G.}~\bibnamefont {Wlaz{\l}owski}}, \bibinfo {author} {\bibfnamefont {K.}~\bibnamefont {Sekizawa}}, \bibinfo {author} {\bibfnamefont {P.}~\bibnamefont {Magierski}}, \bibinfo {author} {\bibfnamefont {A.}~\bibnamefont {Bulgac}},\ and\ \bibinfo {author} {\bibfnamefont {M.~M.}\ \bibnamefont {Forbes}},\ }\href {https://link.aps.org/doi/10.1103/PhysRevLett.117.232701} {\bibfield  {journal} {\bibinfo  {journal} {Phys. Rev. Lett.}\ }\textbf {\bibinfo {volume} {117}},\ \bibinfo {pages} {232701} (\bibinfo {year} {2016})}\BibitemShut {NoStop}%
\bibitem [{\citenamefont {Okihashi}\ and\ \citenamefont {Matsuo}(2021)}]{okihashi2021}%
  \BibitemOpen
  \bibfield  {author} {\bibinfo {author} {\bibfnamefont {T.}~\bibnamefont {Okihashi}}\ and\ \bibinfo {author} {\bibfnamefont {M.}~\bibnamefont {Matsuo}},\ }\href {https://academic.oup.com/ptep/article/doi/10.1093/ptep/ptaa174/6020276} {\bibfield  {journal} {\bibinfo  {journal} {Prog. Theor. Exp. Phys.}\ }\textbf {\bibinfo {volume} {2021}},\ \bibinfo {pages} {023D03} (\bibinfo {year} {2021})}\BibitemShut {NoStop}%
\bibitem [{\citenamefont {Sumiyoshi}\ \emph {et~al.}(2022)\citenamefont {Sumiyoshi}, \citenamefont {Furusawa}, \citenamefont {Nagakura}, \citenamefont {Harada}, \citenamefont {Togashi}, \citenamefont {Nakazato},\ and\ \citenamefont {Suzuki}}]{sumiyoshi2023}%
  \BibitemOpen
  \bibfield  {author} {\bibinfo {author} {\bibfnamefont {K.}~\bibnamefont {Sumiyoshi}}, \bibinfo {author} {\bibfnamefont {S.}~\bibnamefont {Furusawa}}, \bibinfo {author} {\bibfnamefont {H.}~\bibnamefont {Nagakura}}, \bibinfo {author} {\bibfnamefont {A.}~\bibnamefont {Harada}}, \bibinfo {author} {\bibfnamefont {H.}~\bibnamefont {Togashi}}, \bibinfo {author} {\bibfnamefont {K.}~\bibnamefont {Nakazato}},\ and\ \bibinfo {author} {\bibfnamefont {H.}~\bibnamefont {Suzuki}},\ }\href@noop {} {\bibfield  {journal} {\bibinfo  {journal} {Progress of Theoretical and Experimental Physics}\ }\textbf {\bibinfo {volume} {2023}},\ \bibinfo {pages} {013E02} (\bibinfo {year} {2022})}\BibitemShut {NoStop}%
\bibitem [{\citenamefont {Hinohara}\ \emph {et~al.}(2024)\citenamefont {Hinohara}, \citenamefont {Oishi},\ and\ \citenamefont {Yoshida}}]{hinohara2024}%
  \BibitemOpen
  \bibfield  {author} {\bibinfo {author} {\bibfnamefont {N.}~\bibnamefont {Hinohara}}, \bibinfo {author} {\bibfnamefont {T.}~\bibnamefont {Oishi}},\ and\ \bibinfo {author} {\bibfnamefont {K.}~\bibnamefont {Yoshida}},\ }\href {https://link.aps.org/doi/10.1103/PhysRevC.109.034302} {\bibfield  {journal} {\bibinfo  {journal} {Phys. Rev. C}\ }\textbf {\bibinfo {volume} {109}},\ \bibinfo {pages} {034302} (\bibinfo {year} {2024})}\BibitemShut {NoStop}%
\bibitem [{\citenamefont {Hayami}\ \emph {et~al.}(2021)\citenamefont {Hayami}, \citenamefont {Okubo},\ and\ \citenamefont {Motome}}]{hayami2021}%
  \BibitemOpen
  \bibfield  {author} {\bibinfo {author} {\bibfnamefont {S.}~\bibnamefont {Hayami}}, \bibinfo {author} {\bibfnamefont {T.}~\bibnamefont {Okubo}},\ and\ \bibinfo {author} {\bibfnamefont {Y.}~\bibnamefont {Motome}},\ }\href {https://www.nature.com/articles/s41467-021-27083-0} {\bibfield  {journal} {\bibinfo  {journal} {Nat Commun}\ }\textbf {\bibinfo {volume} {12}},\ \bibinfo {pages} {6927} (\bibinfo {year} {2021})}\BibitemShut {NoStop}%
\bibitem [{\citenamefont {Lee}\ and\ \citenamefont {Mochizuki}(2022)}]{lee2022}%
  \BibitemOpen
  \bibfield  {author} {\bibinfo {author} {\bibfnamefont {M.-K.}\ \bibnamefont {Lee}}\ and\ \bibinfo {author} {\bibfnamefont {M.}~\bibnamefont {Mochizuki}},\ }\href {https://link.aps.org/doi/10.1103/PhysRevApplied.18.014074} {\bibfield  {journal} {\bibinfo  {journal} {Phys. Rev. Appl.}\ }\textbf {\bibinfo {volume} {18}},\ \bibinfo {pages} {014074} (\bibinfo {year} {2022})}\BibitemShut {NoStop}%
\end{thebibliography}%

\end{document}